\documentclass[11pt]{article}
\usepackage{graphicx, amsmath, amsfonts, amssymb, amsthm, mathrsfs} % Required 
\usepackage[setpagesize=false,pagebackref=false, 
linktocpage, bookmarksopen=true, colorlinks=true, 
linkcolor=blue,citecolor=blue,urlcolor=blue]{hyperref}
\usepackage{color}
\usepackage{dynkin-diagrams}
\usepackage{booktabs}
\usepackage{tikz}
\usetikzlibrary{arrows.meta, positioning, calc, decorations.pathmorphing, shapes.geometric}
\usepackage{caption}
\newlength{\floatspacing}
\usepackage{latexsym,amsmath,amssymb,theorem,epsfig}
\usepackage[margin=1in]{geometry}

\newcommand{\SLZ}{\ensuremath{\mathrm{SL}(2,\mathbb{Z})}}
\newcommand{\SLR}{\ensuremath{\mathrm{SL}(2,\mathbb{R})}}

\title{\bfseries A 12d Origin of the Type IIB String Theory: A Review}
\author{Chang Cheng\\[2pt]
\normalsize Department of Physics, Imperial College London, SW7 2AZ, U.K.\\
\normalsize \texttt{chang.cheng24@imperial.ac.uk}}
\date{\today}

\begin{document}

\maketitle

\begin{abstract}
\noindent
Type IIB string theory possesses a conjectured non-perturbative \SLZ{} S-duality that is not realised geometrically in ten dimensions (10d).
We examine the proposal that this duality has a twelve-dimensional (12d) origin.
In the axio-dilaton sector the picture is robust: the D7 brane and D(-1) instanton arise as a Kaluza-Klein monopole and pp-wave of 12d gravity reduced on a torus, on which the \SLZ{} acts geometrically as large diffeomorphisms.
Beyond this sector it weakens to a covariant repackaging of the 10d theory: the form fields and their dualities admit a 12d-covariant organisation, and the D3-brane worldvolume duality ties to the bulk \SLZ{}, but a full 12d uplift is obstructed.
In particular, we identify a U(1)-preserving five-point coupling in the higher-derivative effective action that admits no 12d-covariant form.
The 12d picture thus remains an effective organisation of the 10d theory rather than a fundamental spacetime.
\end{abstract}

\tableofcontents
\clearpage
\section{Introduction}
\begin{figure}[ht]
\centering
\includegraphics[width=0.3\textwidth]{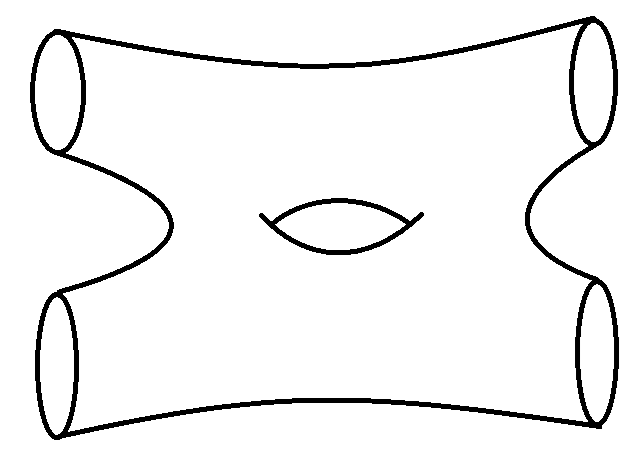}
\caption{A closed-string scattering amplitude is a smooth two-dimensional worldsheet. Where the Feynman diagrams of point particles meet at singular interaction vertices, string interactions are carried by the smooth topology of the worldsheet.}
\label{fig:worldsheet}
\end{figure}

String theory is a framework that attempts to unify the two pillars of fundamental physics: quantum mechanics and general relativity.
It does so by replacing point-like particles with one-dimensional strings in the underlying description.
The singular interaction vertices of point-particle Feynman diagrams give way to smooth two-dimensional worldsheets (Figure~\ref{fig:worldsheet}).
It is known to be consistent in 10 (for supersymmetric strings) and 26 (for bosonic strings) dimensions.
Our 4-dimensional spacetime is then understood as the compactification of a higher-dimensional theory on an internal manifold,
\[
\text{10d spacetime} = \underbrace{\mathbb{R}^{3,1}}_{\text{observed 4d}} \times \underbrace{M_6}_{\substack{\text{compact 6d} \\ \text{(curled up)}}}.
\]

\begin{figure}[ht]
\centering
\includegraphics[width=0.7\textwidth]{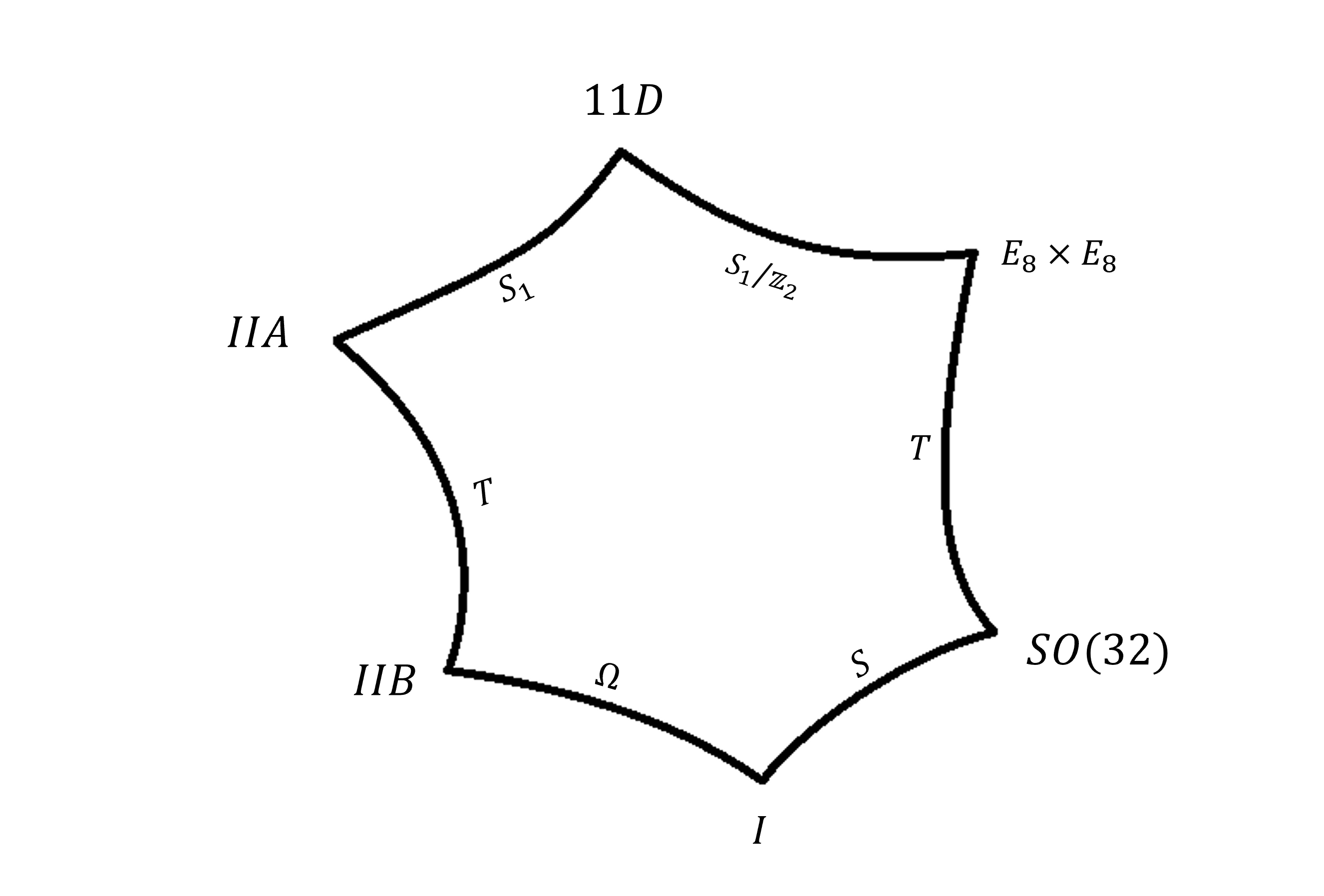}
\caption{The web of superstring dualities and their common origin in eleven-dimensional M-theory. Type IIA and heterotic $E_8\times E_8$ arise from M-theory on a circle ($S^1$) and on an interval ($S^1/\mathbb{Z}_2$). Type IIB arises as the decompactifying limit of M-theory on a two-torus ($T^2$), the nine-dimensional detour that we seek to bypass. Type I follows from a type IIB orientifold ($\Omega$) and is S-dual to heterotic SO(32), while the T-duality links relate the two type II and the two heterotic theories.}
\label{fig:duality-web}
\end{figure}

In 10 dimensions, there are five consistent superstring theories: type IIA, type IIB, heterotic SO(32), heterotic $E_8 \times E_8$, and type I.
These theories are related by dualities, and are believed to be different limits of a single underlying theory, M-theory, in 11 dimensions (Figure~\ref{fig:duality-web}).
For instance, the type IIA and the heterotic $E_8 \times E_8$ theories are understood to arise from M-theory compactified on a circle and an interval, respectively.
The type IIB theory, in turn, arises as the decompactifying limit of the 9d theory obtained from M-theory on a torus.
The type I theory can be obtained from the type IIB theory by an orientifold projection (gauging worldsheet parity), and is related to the heterotic SO(32) theory by S-duality.

An intriguing feature of the type IIB theory is its conjectured \SLZ{} duality, which exchanges the fundamental string with the D1 brane.
This duality is believed to be an exact non-perturbative symmetry of the type IIB theory, and has received support from modular-invariant $R^4$ corrections to the effective action, as well as the existence of the (p, q) strings and 7-branes.
But why does the type IIB theory have such an \SLZ{} duality?
One may reconcile this with the M-theory origin of the type IIB theory, since \SLZ{} is the group of large diffeomorphisms of the torus, whose modulus can be identified with the axio-dilaton.
Yet this explanation is unsatisfactory, as the \SLZ{} duality of the type IIB theory is a fundamental feature in 10d, and thus one should not have to take a detour through 9d to understand it.
This motivates a higher-dimensional account in which the duality is manifest from the outset, without singling out a compactification circle: a 12d origin of the type IIB theory.

Indeed, after the construction of the 11d supergravity \cite{Nahm:1977tg,Cremmer1978}, it was soon realised that SO(10, 2) Majorana-Weyl spinors might offer a natural setting for a 12d supersymmetric theory of gravity \cite{Castellani:1982ke,Blencowe:1988sk}.
Later, through the study of the 6d (2,0) gravity theory \cite{Hull:1995xh,Witten:1995em}, the 7-brane backreacted vacua \cite{Vafa:1996xn}, and the Euclidean D(-1) instanton \cite{Tseytlin:1996it}, more evidence for an effective 12d interpretation of the type IIB theory emerged.

However, the construction of a 10+2d supergravity \cite{Nishino:1997gq} has faced persistent conceptual obstructions, due to its non-compact little group and the lack of a momentum generator in its algebra \cite{Hewson:1997wv}.
To this day, no consistent 10+2d supergravity with 32 supercharges that reduces to the type IIB supergravity has been constructed. 
On the other hand, effective 12d perspectives, arising from branes \cite{Tseytlin:1996ne}, dualities \cite{Tseytlin:1996it}, and effective actions \cite{Minasian:2015bxa}, have proven more fruitful, providing concrete insights into the type IIB theory and its \SLZ{} duality.
These will be reviewed in this paper.\footnote{
One may argue that the appearance of \SLZ{} is merely a manifestation of redundancy in the parameterisation we have chosen to describe the type IIB theory.
While it would be interesting to ask whether a formulation without this redundancy exists, and what it would imply for the other superstring theories and the web of dualities as a whole, that is not our focus here.
As long as no such formulation is known, the \SLZ{}-manifest description remains the one that reveals the web of dualities, and hence is the most natural one we have access to at the moment.
}

Beyond these formal motivations, a new reason to revisit the question comes from holography, specifically the AdS/CFT correspondence.
It appears most concretely in the duality between M-theory and type IIA, where the near-horizon geometry of a stack of M2-branes, $\mathrm{AdS}_4 \times S^7/\mathbb{Z}_k$, is holographically dual to the ABJM theory \cite{Aharony:2008ug}.
The $\tfrac{1}{2}$-BPS circular Wilson loop of this theory admits a dual description as a wrapped M2-brane, whose action reproduces the Wilson-loop expectation value at large $N$ \cite{Giombi:2023vzu,Tseytlin:2024euk}.
The ABJM theory has no continuous coupling, but its discrete Chern-Simons level $k$ may be effectively treated as a coupling, and here it is geometrised as the size of the M-theory circle.
It is natural to ask whether an analogous relation holds on the type IIB side, between the $\mathrm{AdS}_5 \times S^5$ background and the $\mathcal{N}=4$ SYM theory, which carries its own \SLZ{} duality acting on its complexified coupling.
If the $\mathrm{AdS}_5 \times S^5$ background, like the D(-1) and D7 backgrounds, admits a 12d uplift, then wrapping a 12d object on a torus would geometrise the coupling of $\mathcal{N}=4$ SYM.
Addressing this question lies beyond the scope of this paper, and would require a 12d interpretation of the D3 brane that sources the $\mathrm{AdS}_5 \times S^5$ background.

We therefore revisit the question of a 12d interpretation of the type IIB string theory, with additional motivation from the recent developments in AdS/CFT.
This work is motivated by two guiding questions: whether the type IIB theory has a higher-dimensional origin, and what the role and origin of its \SLZ{} duality are.
We pursue them along four lines:
\begin{enumerate}
    \item to establish a 12d interpretation of the type IIB axio-dilaton sector, by realising the D7 brane and D(-1) instanton backgrounds as a 12d Kaluza-Klein monopole and pp-wave, respectively, reduced on a torus, and to verify its consistency with supersymmetry (Section~\ref{sec:axio-dilaton-12d}).
    \item to determine whether this interpretation extends beyond the axio-dilaton sector, by constructing a covariant 12d unification of the type IIB form fields (Section~\ref{sec:covariant-unification}).
    \item to clarify the role of the D3 brane in the type IIB S-duality, by examining how its worldvolume electromagnetic duality is tied to the bulk \SLZ{} S-duality (Section~\ref{sec:d3-self-duality}).
    \item to test the 12d interpretation with the higher-derivative corrections that appear to render \SLZ{} exact, and to clarify its limitations (Section~\ref{sec:effective-actions}).
\end{enumerate}

This paper is organised as follows. Section~\ref{sec:iib-puzzles} introduces the type IIB theory and its puzzles, and Section~\ref{sec:12d-supergravity-attempts} reviews, for completeness, earlier works relevant to the construction of a 10+2d supergravity. 
The four objectives above are then taken up in turn in Sections~\ref{sec:axio-dilaton-12d}--\ref{sec:effective-actions}, with the analogous type IIA constructions presented throughout. We conclude in Section~\ref{sec:outlook} with a summary and open directions.
\section{Type IIB and its Puzzles}
\label{sec:iib-puzzles}
We set the stage for the type IIB supergravity by writing down its spectrum, action, and puzzles.
Early on, it was understood that to have a unitary, interacting theory of gravity, the particles can have at most spin 2 \cite{Weinberg:1964ew,Weinberg:1965nx}, which limits the number of supercharges to at most 32 \cite{Nahm:1977tg}. 
The most notable theories with 32 supercharges are the 11d, type IIA, and type IIB supergravities. 
These theories are believed to be the low-energy limits of the corresponding membrane and string theories \cite{Green:2012oqa}.

Einstein's theory of gravity is a nonlinear theory with diffeomorphism symmetry.
However, one can expand it to quadratic order around a flat spacetime with $g_{mn} = \eta_{mn} + \kappa h_{mn}$, where $\kappa$ is the gravitational coupling.
Then the Einstein-Hilbert action becomes that of a free massless spin-2 field $h_{mn}$, with a rigid Lorentz symmetry SO(d-1, 1), as well as possibly a rigid R-symmetry acting on the supercharges.
On-shell single-particle (free) states furnish representations of the little group SO(d-2), from which one can determine the theory's low-energy spectrum.

The bound of 32 supercharges implies that a supergravity theory can have at most 128 bosonic and 128 fermionic on-shell degrees of freedom.
In 11d, they transform under irreducible representations of SO(9), and consist of a gravitino, a graviton, and a 3-form potential.
The 11d supergravity multiplet can be given with SO(9) Dynkin labels as\footnote{Our convention for Dynkin Labels is given in Appendix~\ref{sec:Dynkin Label Conventions}.}
\begin{equation}
G_{11} = [2000]_{SO(9)} + [0010]_{SO(9)} + [1001]_{SO(9)}.
\label{11d SUGRA multiplet}
\end{equation}
This can be reduced to 10d by decomposing SO(9) representations into those of SO(8).\footnote{
Specifically, the decomposition is given by branching rules of the form
\begin{equation}
\begin{aligned}
[2000]_{SO(9)} &= [2000]_{SO(8)} + [1000]_{SO(8)}+[0000]_{SO(8)}\\
[0010]_{SO(9)} &= [0011]_{SO(8)} + [0100]_{SO(8)}\\
[1001]_{SO(9)} &= [1001]_{SO(8)} + [1010]_{SO(8)}+[0010]_{SO(8)}+[0001]_{SO(8)}.
\end{aligned}
\end{equation}
}
One finds the type IIA multiplet (with SO(8) Dynkin labels)
\begin{equation}
\begin{aligned}
G_{IIA} 
&= \underbrace{[2000]+[0100]+[0000]}_{NSNS} + \underbrace{[0011] + [1000]}_{RR} + [1001] + [0010] + [1010] + [0001]\\
&=([1000] + [0001]) \times ([1000] + [0010]).
\end{aligned}
\label{IIA multiplet}
\end{equation}
Remarkably, the type IIA multiplet \eqref{IIA multiplet} factorises as the product of two vector multiplets of different chirality in 10d.
In this product, the bosonic fields are classified into NSNS and RR sectors: the NSNS sector is that obtained from the tensor product of two vectors, and the RR sector is that obtained from two spinors \cite{Polchinski:1998rr}.
We note that the NSNS/RR distinction is not apparent when the IIA multiplet is viewed as dimensionally reduced from 11d supergravity, but only becomes apparent when viewed as 10d tensor products.
This is related to how membranes live naturally in 11d \cite{Bergshoeff:1987cm}, while strings live in 10d \cite{Green:2012oqa}, and the NSNS/RR classification has a stringy origin \cite{Polchinski:1995mt}.

The factorisation of the type IIA multiplet \eqref{IIA multiplet} motivates the construction of the other supergravity multiplet with 32 supercharges, obtained by taking the tensor product of two vector multiplets of the same chirality. This is the (chiral) type IIB supergravity multiplet
\begin{equation}
\begin{aligned}
G_{IIB}
&=\left([1000] + [0001]\right)^2\\
&=\underbrace{[2000]+[0100] + [0000]}_{NSNS} 
+ \underbrace{[0002] + [0100] + [0000]}_{RR} + 2\cdot[1001] + 2\cdot[0010].
\end{aligned}
\end{equation}
Unlike the type IIA supergravity theory, the type IIB theory in 10 dimensions is not known to follow from the dimensional reduction of another theory.
This is the first puzzle of the type IIB theory: does it have a higher-dimensional origin?

We have been discussing the type IIB supergravity multiplet, which captures the massless spectrum of the quadratic (free) theory with SO(9, 1) symmetry and a rigid SO(2) $\cong$ U(1) symmetry acting on the supercharges.
The nonlinear (interacting) theory is obtained by restoring an infinite number of higher-order terms in the gravitational coupling $\kappa$, which promote the rigid SO(9,1) to diffeomorphisms%
\footnote{
For example, promoting $\frac{1}{2}\int d^{10}x\left(\frac{1}{2}\partial_p h_{mn}\partial^p h^{mn}-\frac{1}{4}\partial_m h \partial^m h\right)$ in de Donder gauge to $\frac{1}{2\kappa^2}\int d^{10}x\sqrt{-g}R$}
and rigid U(1) to local U(1). Doing so \cite{Green:1982tk,Schwarz:1983wa,Schwarz:1983qr,Howe:1983sra}, we find that the theory has a hidden global symmetry group \SLR{}, which, when combined with the now localised U(1) symmetry, gives $\frac{\SLR{}}{U(1)}$ as the scalar manifold. The construction of the coset representative $\mathcal{V}$ is reviewed in Appendix~\ref{sec:sl2r-coset}.

\subsection{The Type IIB Action in Various Frames}

The action of the type IIB supergravity \cite{Schwarz:1983qr,Howe:1983sra,Bergshoeff:1995as} with the string frame metric $G_{mn}$ can be written as 
\cite{Becker:2006dvp,ValeixoBento:2025emh}%
\footnote{
We will not follow the PST approach \cite{Pasti:1996vs,DallAgata:1997gnw,DallAgata:1998ahf}, and instead impose self-duality as an additional field equation.
}
\begin{equation}
\begin{aligned}
S_{IIB}
&=S_{NSNS} + S_{RR} + S_{CS}\\
&=\frac{1}{2\kappa^2}
\int d^{10}x \sqrt{-G}
\Bigg[
e^{-2\Phi}\left(R + 4(\partial\Phi)^2-\frac{1}{2}|H_3|^2\right)
-\left(\frac{1}{2}(\partial C_0)^2+\frac{1}{2}|\tilde F_3|^2+\frac{1}{4}|\tilde{F}_5|^2\right)
\Bigg]\\
&\quad -\frac{1}{4\kappa^2}\int C_4\wedge H_3\wedge F_3
\end{aligned}
\label{string frame IIB action}
\end{equation}
with $2\kappa^2=(2\pi)^7l_s^8=(2\pi)^7\alpha^{\prime 4}$, and
\begin{equation}
F_p = dC_{p-1}, \quad 
H_3 = dB_2, \quad
\tilde F_3 = F_3 - C_0 H_3,\quad
\tilde F_5 = * \tilde F_5 = F_5 - \frac12 C_2\wedge H_3 + \frac12 B_2\wedge F_3.
\end{equation}
We note that \eqref{string frame IIB action} is invariant, up to a boundary term, under the \SLR{} transformations
\begin{equation}
G_{mn}\to |c\tau+d|G_{mn},\quad
\tau\equiv C_0 + i e^{-\Phi} \to \frac{a\tau+b}{c\tau+d},\quad
\begin{pmatrix}
H_3\\F_3
\end{pmatrix}
\to
\begin{pmatrix}
d&c\\b&a
\end{pmatrix}
\begin{pmatrix}
H_3\\F_3
\end{pmatrix},\quad
F_5 \to F_5.
\label{string frame sl2r transformations}
\end{equation}
The action of \SLR{} on the metric in \eqref{string frame sl2r transformations} arises because the string frame Einstein-Hilbert term carries an explicit dilaton coupling.
It is therefore natural to pass to an Einstein frame metric \cite{Becker:2006dvp,Weigand:2018rez}, in which the Einstein-Hilbert action takes its canonical form.
This condition fixes the Einstein frame metric only up to an overall conventional factor, leaving two common but mutually exclusive choices:
\begin{enumerate}
    \item that the Einstein frame metric be \SLR{}-invariant, or
    \item that the Einstein frame metric be asymptotically flat.
\end{enumerate}
The \SLR{}-invariant Einstein frame metric is given by
\begin{equation}
g_{mn} \equiv e^{-\Phi/2}G_{mn}.
\label{sl2r invariant Einstein frame metric}
\end{equation}
From string perturbation theory, the string frame metric $G_{mn}$ is taken to be asymptotically flat: $G_{mn} \to \eta_{mn}$. Hence the \SLR{}-invariant Einstein frame metric is not: $g_{mn} \to g_s^{-1/2}\eta_{mn}$.
There is an implicit difference in length scale between the string and Einstein frames, $l_E = g_s^{-1/4}l_s$, which one must account for when comparing tensions computed in the Einstein and string frames.
The action is
\begin{equation}
\begin{aligned}
S_{IIB}
&=\frac{1}{2\kappa^2} \int d^{10}x\sqrt{-g}\left[\left(R-\frac{\partial\bar\tau\partial\tau}{2\tau_2^2}\right)
-\frac{1}{2}\left(
e^{-\Phi}|H_3|^2+e^\Phi|\tilde{F}_3|^2
\right)
-\frac{1}{4}|\tilde F_5|^2
\right]\\
&-\frac{1}{4\kappa^2}\int C_4\wedge H_3\wedge F_3.
\end{aligned}
\label{eq:sl2r_invariant_einstein_frame_action}
\end{equation}
Had one taken the second route and worked with an asymptotically-flat Einstein frame metric $\tilde{g}_{mn} \equiv g_s^{1/2}e^{-\Phi/2}G_{mn}$, there would no longer be a difference in length scale between the two frames, but the metric would not be \SLR{}-invariant:
\begin{equation}
\tilde{g}_{mn} \to |c\langle\tau\rangle + d| \tilde{g}_{mn}.
\end{equation}
Hence to write down an \SLR{}-invariant action, appropriate factors of $g_s$ must be included. For example, while $R(\tilde{g}_{mn})$ and $\sqrt{-\tilde{g}} R(\tilde{g}_{mn})$ are not \SLR{}-invariant, 
$g_s^{1/2}R(\tilde{g}_{mn})$ and $\frac{1}{g_s^2}\sqrt{-\tilde{g}} R(\tilde{g}_{mn})$ are invariant.
Using the asymptotically-flat Einstein frame metric $\tilde{g}_{mn} = g_s^{1/2}e^{-\Phi/2}G_{mn}$, the type IIB action is
\begin{equation}
\begin{aligned}
S_{IIB}
&=\frac{1}{2\kappa^2} \int d^{10}x\sqrt{-\tilde{g}}\left[
\frac{1}{g_s^2}\left(R-\frac{\partial\bar\tau\partial\tau}{2\tau_2^2}\right)
-\frac{1}{2 g_s}\left(
e^{-\Phi}|H_3|^2+e^\Phi|\tilde{F}_3|^2
\right)
-\frac{1}{4}|\tilde F_5|^2
\right]\\
&-\frac{1}{4\kappa^2}\int C_4\wedge H_3\wedge F_3,
\end{aligned}
\label{eq:asymptotically_flat_einstein_frame_action}
\end{equation}
where the factors of $g_s$ necessarily enter in order to ensure \SLR{}-invariance.
In this convention, it is also standard to define the gravitational coupling $\kappa_{10}^2 = g_s^2 \kappa^2$.
Because we will primarily be concerned with the \SLR{} symmetry, it is more convenient to work with the \SLR{}-invariant Einstein frame metric \eqref{sl2r invariant Einstein frame metric}, where explicit factors of $g_s$ do not enter.

Although the scalars $\Phi, C_0$, and the form fields $C_2, B_2$ have different NSNS and RR origins in 10d, they transform into each other under \SLR{}.
This \SLR{}-invariance will hold at the two-derivative level, as well as for higher-derivative corrections with trivial dependence on $\tau$.
Once corrections with nontrivial $\tau$-dependence enter, the \SLR{} symmetry is broken.
As string and brane corrections are included, the symmetry of \SLZ{} may be restored, which is understood as a reflection of the quantisation of NSNS and RR charges in type IIB string theory.

\subsection{Guiding Questions}

As discussed above, the NSNS/RR distinction is a purely 10d one.
Both the NSNS and RR fields in type IIA combine to form SO(9) multiplets.
Analogously, the NSNS and RR fields in type IIB combine to form $\SLR{}\times SO(8)$ multiplets, and it is natural to ask whether this could be a hint of a higher-dimensional origin.
This is the second puzzle of the type IIB theory, and it is unlikely to be independent of the first: where does the \SLZ{}\footnote{
\SLR{} at the two-derivative supergravity level.
} symmetry come from?\footnote{
The type IIB theory is understood as the decompactifying limit of M-theory on a torus, which could explain \SLZ{}. But \SLZ{} is a true symmetry in 10d already.}
We are motivated by the two guiding questions introduced above:
\begin{enumerate}
    \item Does the type IIB theory have a higher-dimensional origin?
    \item What are the role and origin of \SLZ{}?
\end{enumerate}
\section{A 10+2d Theory?}\label{sec:12d-supergravity-attempts}
There exists an algebra that admits a 10+2d interpretation and may unify the algebras of the various theories with 32 supercharges. 
Historically, this motivated a series of investigations into formulating a supersymmetric 10+2d theory of gravity.
These constructions will be reviewed for completeness.
The objective here is to present these attempts in an organised fashion, discuss their obstructions, and set the stage for the brane and effective action discussions later.

In 12d the complex Dirac spinor has 64 components and decomposes into two 32-component Weyl spinors.
In general, they are complex, but for the Lorentz group SO(10, 2),\footnote{We write SO(s, t) for the orthogonal group of a metric with $s$ spacelike and $t$ timelike directions, so SO(10, 2) has signature $(10, 2)$.} because
\begin{equation}
s-t\equiv 0 \pmod 8,
\end{equation}
one can impose a Majorana condition compatible with chirality to find Majorana-Weyl spinors with 32 real components \cite{VanProeyen:1999ni}.
The corresponding superalgebra contains a 2-form and a self-dual 6-form charge
\cite{vanHolten:1982mx}
\begin{equation}
\operatorname{Sym}^2[000001]_{SO(10, 2)} = [010000]_{SO(10, 2)} + [000020]_{SO(10, 2)}.
\label{OSp algebra}
\end{equation}
This algebra is often referred to as the $\mathrm{OSp}(1|32)$ algebra \cite{Bergshoeff:2000qu} or the F-theory algebra \cite{Hewson:1999tf}.
It reduces to the 11d algebra
\begin{equation}
\operatorname{Sym}^2[00001]_{SO(10, 1)} = [10000]_{SO(10, 1)} + [01000]_{SO(10, 1)} + [00002]_{SO(10, 1)}
\end{equation}
and the 10d type IIA and type IIB algebras \cite{Bergshoeff:2000qu,Bergshoeff:2000vg,Bergshoeff:2000vh}.
This unifying role made it a natural starting point for attempts to formulate a supersymmetric 10+2d theory of gravity \cite{Castellani:1982ke}, and for studying the objects such a theory would contain. 
For instance, 2+2d branes can propagate in (10, 2) spacetime \cite{Blencowe:1988sk}. 
Their BPS states \cite{Ueno:1999xa} and the fractions of supersymmetry they preserve \cite{Manvelyan:1998kx} have been classified. 
At a purely algebraic level, these ideas extend to proposals for unifying the various dualities in 12 and 13 dimensions \cite{Bars:1996dz,Bars:1996us,Bars:1996ab,Bars:1996hr}, and even in 14 dimensions \cite{Rudychev:1997ui,Bars:1997ug}.

However, the possibility of a 10+2d theory hinges on there being two time-like directions, and there are fundamental obstacles to formulating a supersymmetric theory with two times.
The little group of an SO(10, 2) theory is the non-compact SO(9, 1), with no nontrivial unitary irreducible representations \cite{Weinberg:1995mt}.
For unitary representations of the supersymmetry algebra, we may write the supercharge anticommutator 
\begin{equation}
\{Q_\alpha,Q_\beta^\dagger\}=\sum_n(\Gamma_n)_{\alpha\beta} Z_n,
\end{equation}
where $\Gamma_n$ are the antisymmetrised gamma-matrix structures and $Z_n$ the central charges. 
Because the RHS is symmetric and positive definite, it can be diagonalised.
Then we obtain fermionic raising and lowering operators, which implies that there are 256 states in the system. 
However, this cannot be carried out if the representation is non-unitary.
One might say, in any known one-time supergravity theory with 32 supercharges, there are 128+128 states, so start with this many as well. 
But the gravitino representation of SO(10) already has 144 states, exceeding the budget for fermions. There indeed is a way to add up to 144 + 144 states for SO(10) without too many scalars: a single gravitino in the fermion sector plus a graviton and two 2-forms in the bosonic sector. 
But this does not seem related to any known 10d or 11d multiplets.

Another issue that accompanies a non-compact little group is negative-norm states, or ``ghosts''.
Partially motivated by studying the 10+2d theory, a practical two-time framework has been developed \cite{Bars:1998cs,Bars:1998ui,Bars:1999nk,Bars:1999uq,Bars:2000qm}. 
The key ingredient is a local Sp(2, $\mathbb{R}$) gauge symmetry acting on the phase-space variables, treated as a two-component vector.
Different gauge choices lead to different one-time systems that share a common (d-2, 2) parent description.
The various approaches to formulating 10+2d theories largely follow this logic.

There have also been attempts at 10+2d SYM \cite{Nishino:1996wp,Sezgin:1997gr,Nishino:1997hk,Nishino:1997ia,Nishino:1999ck} and 10+2d supergravity \cite{Nishino:1997gq,Nishino:1997sw,Nishino:1998qn}, but none of them have achieved satisfactory results, as they either introduce null projectors that explicitly break SO(10, 2),
or fail to construct vielbeins due to the lack of a momentum generator \cite{Hewson:1999tf} in the algebra \eqref{OSp algebra}. 

After it was shown that the 2+2d brane can propagate in 10+2d spacetime \cite{Blencowe:1988sk}, the $\mathcal{N}=(2,1)$ string was also studied \cite{Kutasov:1996fp,Kutasov:1996zm,Martinec:1996wn}. The $\mathcal{N}=1$ sector contains an effective 10+2d target space, with the $\mathcal{N}=2$ sector on a 2+2d target.
Later, an $\mathcal{N}=1$ superstring with a 2+2d target space was constructed \cite{Khviengia:1995pm,Lu:1995pn}.
Further work investigated whether self-dual gravity in 2+2 dimensions admits a stringy description \cite{Sezgin:1995fj} and related these self-dual 2+2d strings to a supersymmetric membrane action with $\mathrm{OSp}(1|32)$ and $\mathrm{OSp}(8|2)$ subgroup structure \cite{Ketov:1996gr}.
In parallel, a Green-Schwarz-type super 2+2d brane embedded in a 10+2d background was constructed \cite{Hewson:1996yh,Hewson:1997wv,Hewson:1998sw}.

There were also investigations into higher-dimensional bosonic field theories that, upon compactification and truncation, may produce the known theories.
In \cite{Khviengia:1997rh}, a 12d action with imaginary dilaton couplings was suggested as
\begin{equation}
2\kappa_{12}^2S
=\int d^{12}x\sqrt{-\widehat{g}}
\left[
\widehat R-\frac{1}{2}(\partial\widehat{\Phi})^2
-\frac{1}{2}e^{\frac{2i}{\sqrt{5}}\widehat{\Phi}}|\widehat{F}_5|^2
-\frac{1}{2}e^{\frac{i}{\sqrt{5}}\widehat{\Phi}}|\widehat{F}_4|^2
\right]
+\frac{\sqrt{3}}{4}\int \widehat{C}_4\wedge \widehat{F}_4 \wedge \widehat{F}_4
\label{Khviengia action}
\end{equation}
where $\widehat{F}_5 = d\widehat{C}_4$, $\widehat{F}_4 = d\widehat{C}_3$, $[\kappa_{12}^2]=L^{10}$, and $\widehat{\Phi}$ here is a 12d dilaton-like scalar.
We will use hats $\widehat{\phantom{x}}$ for higher-dimensional fields and straight letters $g_{mn},C_n,F_{n+1}$ for lower-dimensional fields. 
This will be discussed more clearly in Section~\ref{sec:covariant-unification}.
This action \eqref{Khviengia action} was subsequently studied by \cite{Bakhtiarizadeh:2018upg} at higher derivatives.
The most prominent feature of the proposal \eqref{Khviengia action} by \cite{Khviengia:1997rh} is the 12d dilaton and its imaginary couplings.
It was introduced based on a certain scalar invariant \cite{Lu:1995cs} across 11d and 10d type IIB brane solutions \cite{Khviengia:1997rh}.

More recently, it has been claimed that F-theory \cite{Vafa:1996xn} admits a 12d action \cite{Choi:2014vya,Choi:2015gia}, taking the form
\begin{equation}
2\kappa_{12}^2S = \int d^{11}x dy \sqrt{-\widehat{g}}
\left(
\widehat{R}-\frac{1}{2}|\widehat{F}_5|^2
\right)
+\frac{1}{6}
\int \widehat{C}_4\wedge \widehat{F}_4 \wedge \widehat{F}_4.
\label{Choi action}
\end{equation}
Here $\widehat{g}$ is the 12d metric and $\widehat{F}_5$ is a 5-form field strength.
All dependence on the twelfth coordinate $y$ is packaged into a scalar $r\equiv r(x^m,y)$, which also enters the metric.
In particular
\begin{equation}
\begin{gathered}
\widehat{C}_4(x,y) \equiv r(x,y) C_3\wedge dy,\quad
\widehat{F}_5(x,y) \equiv r(x,y) d C_3\wedge dy,\\
\widehat{g}_{mn} = g_{mn},\quad
\widehat{g}_{my} = 0,\quad
\widehat{g}_{yy} = r^2.
\label{eq:choi-embedding}
\end{gathered}
\end{equation}
Upon closer examination, we find \eqref{Choi action} is 11d supergravity integrated over a spectator dimension.
This can be seen by substituting \eqref{eq:choi-embedding} into \eqref{Choi action} and evaluating $\int d^{11}x dy \sqrt{-\widehat{g}} \widehat{R}$.
The action \eqref{Choi action} reduces to
\begin{equation}
2\kappa_{12}^2 S =
\int d^{11}x \lambda(x)\sqrt{-g}
\left(R-\frac{1}{2}|F_4|^2\right)
+\frac{1}{6}\int \lambda(x) C_3\wedge F_4\wedge F_4
+\text{boundary},\quad
\lambda(x)\equiv \int dy r(x,y).
\end{equation}
This is the bosonic action of 11d supergravity, multiplied by an auxiliary field $\lambda(x)$ whose equation of motion imposes that the 11d bosonic Lagrangian vanishes.

Beyond these attempts at an explicit 12d action, several related frameworks approach the question from other directions.
Exceptional Field Theory, for instance, realises the various string theory dualities through extended coordinates rather than additional spacetime dimensions, and an \SLR{}$\times \mathbb{R}^+$ version has been proposed \cite{Berman:2015rcc,Rudolph:2017vzy}, though it is not itself a 12d theory.
Other 12d viewpoints include proposals in which the type IIB theory is an edge mode of a gapped 12d bulk theory, motivated by a 6d generalisation of the fractional quantum Hall effect \cite{Heckman:2017uxe}, as well as the perspective that 2+2 signature spacetime may be a tool for studying Lorentzian QFT \cite{Heckman:2018mxl,Heckman:2022peq}. 
We also note a recent supergravity construction of a locally supersymmetric $SO(10, 2)$-invariant action in $D=10+2$, of MacDowell-Mansouri type and containing the Einstein-Hilbert term \cite{Castellani:2017vbi}.
The relationship between its physical degrees of freedom and those of the type IIB and 11d supergravity remains to be clarified.

The persistent obstructions reviewed above (the non-compact little group, the absence of a momentum generator in the $\mathrm{OSp}(1|32)$ algebra, and the failure to match known 10d or 11d multiplets) suggest that a dynamical 10+2d supergravity with 32 supercharges reducing to the type IIB theory may not be available.
A more modest strategy is then to ask whether specific solutions of the known 10d theories admit a 12d geometric interpretation, without requiring a full dynamical 12d theory.
This is the approach taken in the following sections.

\section{The Axio-Dilaton Sector and 12d Gravity}
\label{sec:axio-dilaton-12d}
We now turn to the 12d structures suggested by brane solutions.
This higher-dimensional effective description is in parallel with what is known between the type IIA theory and M-theory (Table~\ref{tab:iia-iib-uplift}).
The 12d interpretations of the type IIB D7 and D(-1) backgrounds were first identified in \cite{Vafa:1996xn,Tseytlin:1996ne}. Here we derive them explicitly, alongside the corresponding type IIA constructions, and use them to read off how the type IIB \SLZ{} duality acts geometrically in 12d.

\begin{table}[ht]
\centering
\begin{tabular}{lcc}
\toprule
 & \multicolumn{2}{c}{reduced on a circle or torus to}\\
\cmidrule(l){2-3}
higher-dimensional solution & type IIA (from 11d) & type IIB (from 12d)\\
\midrule
Kaluza-Klein monopole & D6 brane & D7 brane\\
pp-wave               & D0 brane & D(-1) instanton\\
\bottomrule
\end{tabular}
\caption{The parallel that organises this section. 
Two solutions of eleven-dimensional supergravity reduce on the M-theory circle to the type IIA D6 brane and D0 brane. 
The same two geometries in twelve dimensions reduce on a torus to the type IIB D7 brane and D(-1) instanton.}
\label{tab:iia-iib-uplift}
\end{table}
One can truncate the type IIB action \eqref{eq:sl2r_invariant_einstein_frame_action} to the axio-dilaton sector
\begin{equation}
2\kappa^2S=
\int d^{10}x\sqrt{-g}\left[
R-\frac{1}{2}\frac{|\partial \tau|^2}{\tau_2^2}
\right].
\label{IIB axio-dilaton action}
\end{equation}
For supersymmetric solutions, we solve for Killing spinors.
In the type IIB theory, the R-symmetry is local $SO(2)\cong U(1)$, and the Killing spinor equations are \cite{Bergshoeff:2006jj}
\begin{equation}
\begin{aligned}
\delta\lambda
&=\frac{i}{\tau-\bar\tau}(\Gamma^\mu\partial_\mu \bar\tau)\epsilon,\\
\delta\psi_\mu
&=
\left(\nabla_\mu+\frac{i}{2}Q_\mu\right)\epsilon,
\end{aligned}
% \tag{0612072 eq3.1}
\label{9+1 IIB axio-dilaton Killing spinor equations}
\end{equation}
where
\begin{equation}
Q_\mu \equiv -\frac{\partial_\mu \tau_1}{2\tau_2}
\label{def-Qm}
\end{equation}
is the non-dynamical U(1) connection built with $\tau$.
The associated integrability condition is that the total holonomy built with the Levi-Civita connection and the U(1) connection vanishes:
\begin{equation}
\left[\frac{1}{4}R_{\mu\nu ab}\Gamma^{ab}
+ \frac{i}{2}F_{\mu\nu}(Q)\right]
\epsilon = 0,\quad
F_{\mu\nu}(Q)=(\nabla_\mu Q_\nu-\nabla_\nu Q_\mu).
\end{equation}
This can be satisfied in two ways: either they cancel each other (as for the D7 brane) or both contributions vanish (as for the D(-1) instanton).
We now present both solutions.

Among the known BPS branes in the type IIA, IIB, and 11d theories \cite{Stelle:1998xg}, the D7 brane plays a special role: it is co-dimension 2, which means its source profile is logarithmic rather than exhibiting the power-law behaviour $\sim \frac{1}{r^{\tilde{d}-2}}$, and thus is not asymptotically flat.
The general $N \times$ D7 brane solution can be written as \cite{Greene:1989ya,Gibbons:1995vg,Tseytlin:1996ne,Bergshoeff:2006jj}
\begin{equation}
ds_{D7}^2 = -dt^2 + d\vec{x}_7^2 + \text{Im} \tau dw d\bar w,
\label{D7 solution}
\end{equation}
where $\tau = C_0 + i e^{-\Phi}$ is the axio-dilaton, and takes the form
\begin{equation}
j(\tau(w)) = \frac{P_N(w)}{P_{N-1}(w)},
\end{equation}
with $j(\tau)$ the holomorphic modular invariant function, and $P_{N-1}(w)$ a degree $N-1$ polynomial, the roots of which are the locations of the $N-1$ D7 branes, the $N$-th D7 brane being at infinity where $P_N/P_{N-1}$ has a pole.
For the case of a single D7 brane localised at $w_i$ (Figure~\ref{fig:d7-monodromy}), the axio-dilaton profile near $w_i$ is dictated by monodromies to be
\begin{equation}
\tau \sim \frac{1}{2\pi i}\ln (w-w_i)+\text{const}
 = \frac{\text{Arg}(w-w_i)}{2\pi} -i\frac{\ln |w-w_i|}{2\pi}+\text{const}.
 \label{D7 axio-dilaton profile}
\end{equation}

\begin{figure}[ht]
\centering
\begin{tikzpicture}[>=Stealth]
  \draw[->] (-0.3,0) -- (5.2,0) node[right] {\footnotesize $\operatorname{Re} w$};
  \draw[->] (0,-0.3) -- (0,3) node[above] {\footnotesize $\operatorname{Im} w$};
  \draw[decorate, decoration={snake, amplitude=0.6pt, segment length=5pt}, thick]
       (2,1.4) -- (4.9,1.4) node[pos=0.74, above] {\footnotesize branch cut};
  \draw[->, thick, blue] (2,1.4) ++(8:0.8) arc (8:352:0.8);
  \node[blue, font=\footnotesize] at (2,2.55) {$\tau\to\tau+1$};
  \fill (2,1.4) circle (2pt);
  \node[anchor=north east] at (1.9,1.33) {\footnotesize $w_i$};
\end{tikzpicture}
\caption{The axio-dilaton around a single D7 brane in the transverse complex plane $w$. Encircling the brane at $w_i$ once sends $\tau\to\tau+1$ through the monodromy $\tau\sim\frac{1}{2\pi i}\ln(w-w_i)$.
}
\label{fig:d7-monodromy}
\end{figure}

On the other hand, the D(-1) is a solution to the Euclidean type IIB theory \cite{Gibbons:1995vg}.
In Einstein frame, it can be written as
\begin{equation}
ds_{10}^2 = \sum_{i=1}^{10} dx_i^2,\quad
e^\Phi = H, \quad\mathcal C\equiv-iC_0 = H^{-1}-1,\quad
H=1+\frac{Q}{r^8},
\label{D-1 background}
\end{equation}
where $\mathcal{C}$ is the Euclidean axion.

\subsection{D7 and D6 Interpreted as 12d and 11d KK-Monopoles}
We will now show that the D7 solution can be interpreted as a 12d KK-monopole geometry compactified on a torus.
This story is in parallel with the story on the type IIA side, between the D6 brane and an 11d KK-monopole geometry.

The ``Kaluza-Klein monopole (KK-monopole)'' \cite{Gross:1983hb,Sorkin:1983ns} refers to the solution of the Einstein-Hilbert action whose KK reduction yields a magnetic monopole.
Thus it can also be understood as the product of Minkowski space and the Taub-NUT space \cite{Ortin:2015hya}.
In $d$ dimensions, the KK-monopole geometry is given by
\cite{Eguchi:1980jx}
\begin{equation}
ds_d^2 = ds_{1,d-5}^2 + H(\vec{y})d\vec{y}^2 + H(\vec{y})^{-1}(dz + \vec{A}\cdot d\vec{y})^2,
\label{d-dim KK-monopole}
\end{equation}
where $\vec{y}$ are 3d Euclidean coordinates and $z$ is a compact coordinate.
Together they form a 4d space that is an $S_1$ fibration over $\mathbb{R}^3$.
Here $H$ and $\vec{A}$ satisfy the curl and source Einstein equations
\begin{equation}
\vec{\nabla} \times \vec A = \vec\nabla H,\quad
\partial^2 H=-\sum_i q_i \delta^3(\vec{y}-\vec{y}_i).
\label{kk monopole einstein eqs}
\end{equation}

\paragraph{D6 brane from reduction of the 11d KK-monopole}
Specialising to 11d, the KK-monopole is given by
\begin{equation}
ds_{11}^2=ds_{1,6}^2 + H(\vec{y})d\vec{y}^2+H(\vec{y})^{-1}(dz+\vec{A}\cdot d\vec{y})^2.
\label{11d KK-monopole}
\end{equation}
It is co-dimension 3, localised on $S_1\times\mathbb{R}^3$. 
We will recognise this $S_1$ factor as the M-theory circle.
By matching \eqref{11d KK-monopole} with the string frame KK reduction ansatz
\begin{equation}
ds_{11}^2=e^{-2\Phi/3}ds_{10}^2+e^{4\Phi/3}(dz+ \vec{A}\cdot d\vec{y})^2,
\label{string frame KK reduction ansatz}
\end{equation}
we find the 10d metric and dilaton profile of
\begin{equation}
ds_{10,string}^2=H^{-1/2}[-dt^2+d\vec{x}_6^2] + H^{1/2}d\vec{y}^2,\quad
e^\Phi=H^{-3/4},
\end{equation}
which is the D6 solution \cite{Horowitz:1991cd,Townsend:1995kk} (see also Appendix~\ref{sec:catalog-of-branes}), with $F_2=*_3 dH$.
This relation is standard within the web of dualities: the theory accounting for the 11d KK-monopole is 11d supergravity, a well-defined, dynamical, supersymmetric theory.
By contrast, although no dynamical 12d supergravity is known, there exists an analogous correspondence between the 12d KK-monopole and the type IIB D7 brane solution \cite{Tseytlin:1996ne}, which we now discuss.

\paragraph{D7 brane from reduction of the 12d KK-monopole}
We begin with the 12d KK-monopole geometry
\begin{equation}
ds_{12}^2 = ds_{1,7}^2 + H(\vec{w})d\vec{w}^2 + H^{-1}(\vec{w})(dz_1 + \vec{A}\cdot d\vec{w})^2.
\label{12d KK-monopole}
\end{equation}
Here $\vec{w} = (w_1, w_2, w_3)$ is a 3d Euclidean vector.
Let the $w_3$ direction be compactified with radius one: $w_3\sim w_3 + 1$.
The curl and source Einstein equations \eqref{kk monopole einstein eqs} become
\begin{equation}
\begin{pmatrix}
\partial_1 H\\
\partial_2 H\\0
\end{pmatrix}
=\begin{pmatrix}\partial_2 A_3\\
-\partial_1 A_3\\
\partial_1 A_2-\partial_2 A_1
\end{pmatrix},\quad
(\partial_1^2+\partial_2^2)H=-\sum_i q_i \delta^2(\vec{w}-\vec{w}_i).
\label{reduced einstein eq}
\end{equation}
We fix a gauge where $A_1=A_2=0$ by performing the following gauge transformation denoted $T$
\begin{equation}
T:z_1\to z_1+\int dw_1 A_1 (w_1,w_2)+n w_3.
\label{T gauge transformation}
\end{equation}
After accounting for the induced transformations on $\vec{A}$ and $H$, the metric becomes
\begin{equation}
ds_{12}^2 = ds_{1,7}^2 + H(dw_1^2 + dw_2^2) + H dw_3^2 + H^{-1}[dz_1 + A_3 dw_3]^2.
\end{equation}
We now rename the coordinates and fields in the following manner:
\begin{equation}w,\bar w\equiv w_1\pm i w_2,\quad
z_2\equiv - w_3,\quad
\tau \equiv -A_3 + i H.
\end{equation}
Then the 12d KK-monopole metric \eqref{12d KK-monopole}, and the corresponding Einstein equations \eqref{reduced einstein eq} take the form
\begin{equation}
ds_{12}^2 = -dt^2 + d\vec{x}_7^2 + \tau_2 dw d\bar w + \tau_2^{-1}|dz_1 + \tau dz_2|^2,\quad
\bar\partial \tau = 0,\quad
\partial\bar\partial \tau_2 = -\sum_i q_i \delta^2(w,w_i).
\label{12d KK-monopole in complex coordinates}
\end{equation}
Such a 12d metric can equivalently be written as the embedding 
\begin{equation}
\widehat{g}_{MN} = 
\begin{pmatrix}
g_{mn} & 0\\
0 & \widehat{g}_{ab}
\end{pmatrix},\quad
\widehat{g}_{ab} = \frac{1}{\tau_2}\begin{pmatrix}
1 & \tau_1\\
\tau_1 & |\tau|^2
\end{pmatrix},
\end{equation}
and we see that $g_{mn}$ is the 10d metric of the D7 brane solution \eqref{D7 solution}.

We now check the profile of $\tau$.
By examining the curl Einstein equation \eqref{kk monopole einstein eqs}, we see that it imposes holomorphy on $\tau$, which then demands that
\begin{equation}
\tau_1= -\sum_i q_i \frac{1}{2\pi}\text{Arg}(w-w_i) + \text{holomorphic}.
\end{equation}
Meanwhile, the source equation in \eqref{12d KK-monopole in complex coordinates} can be solved with
\begin{equation}
\tau_2=-\frac{1}{2\pi}\sum_i q_i \ln |w-w_i|.
\end{equation}
We find that the $\tau$ profile near a source localised at $w_i$ is indeed the profile of the axio-dilaton near a D7 brane \eqref{D7 axio-dilaton profile}. 
For more general $(p,q)$ branes one would need to select the appropriate \SLZ{} section.

We now examine the $S,T$ generators of type IIB \SLZ{}.
The $T$ gauge transformation given in \eqref{T gauge transformation} is precisely the type IIB \SLZ{}-$T$ transformation on $\tau$, and the type IIB \SLZ{}-$S$ transformation is achieved with a 12d coordinate swap
\begin{equation}
w_3\to z_1, \quad
z_1\to -w_3 .
\end{equation}
We can thus identify the type IIB \SLZ{} duality transformations as large diffeomorphisms in 12d.

\paragraph{\boldmath Geometric Interpretation of the $(p,q)$ 7-brane and \SLZ{}}
\label{par:interpreting-sl2z}

The D7 brane found above is a member of the type IIB $(p,q)$ 7-branes, a family permuted by \SLZ{} \cite{Weigand:2018rez}. 
We now give this family a geometric origin in 12d. The claim is that a $(p,q)$ 7-brane is the 12d KK-monopole on $T^2$ whose Taub-NUT circle is wrapped on the $(p,q)$ cycle of the torus.

Having identified the type IIB \SLZ{} with the 12d large diffeomorphisms $T$ and $S$ above, we now recall how the $T$ and $S$ generators of \SLZ{} arise from the periodic identifications that define a torus. 
A torus has topology $S^1 \times S^1$. 
We take coordinates $\vec{\sigma} \in \mathbb{R}^2$ with the periodic identification $(\sigma^1,\sigma^2)\sim (\sigma^1,\sigma^2) + 2\pi (m,n)$ for integers $m,n$.
Diffeomorphism and Weyl symmetry allow the metric to be brought to the flat form $ds^2 = (d\sigma^{1})^2+(d\sigma^{2})^2$ \cite{Polchinski:1998rq}, at the cost of skewing the periodic identification into a parallelogram
\begin{equation}
(\sigma^1,\sigma^2) \sim (\sigma^1,\sigma^2) + 2\pi (m+nv^1,n v^2),
\end{equation}
for some $\vec{v} \in \mathbb{R}^2$.
These identifications define a 2d lattice, most naturally described in the complex plane.
We package the parallelogram vector $\vec{v}$ as the complex modulus $\tau \equiv v^1 + i v^2$ (with $\text{Im} \tau \neq 0$) and adopt the lattice basis $1, \tau$ in place of the orthonormal $1, i$.
The periodic identification becomes
\begin{equation}
\sigma^1 + \tau \sigma^2 \sim \sigma^1 + \tau \sigma^2 + 2\pi(m + n\tau),
\end{equation}
and the metric is
\begin{equation}
ds^2 = |d\sigma^1 + \tau d\sigma^2|^2.
\end{equation}

The torus modulus $\tau$ is therefore defined only up to automorphisms on the lattice, the transformations that leave the torus geometry unchanged.
There are two such transformations:
\begin{enumerate}
    \item $\tau \to \tau+1$: the basis relabelling $(m, n) \to (m+n,n)$ at fixed geometry.
    \item $\tau \to -1/\tau$: the basis relabelling $(m, n) \to (n,-m)$, followed by a constant rescaling of the coordinates that restores the metric to standard form.
\end{enumerate}

\begin{figure}[ht]
\centering
\begin{tikzpicture}[>=Stealth, scale=1.7, every node/.style={font=\footnotesize}]
  \def\lat{\begin{scope}\clip (-1.2,-0.3) rectangle (2.0,2.1);
      \foreach \m in {-1,0,1,2} \foreach \n in {0,1,2}
        \fill[black!35] (\m+0.35*\n,0.95*\n) circle (1.1pt);\end{scope}}
  % ----- T : tau -> tau+1  (shear) -----
  \begin{scope}
    \lat
    \draw[->,thick,black!55] (0,0)--(1,0) node[below right=-1pt]{$1$};
    \draw[->,very thick,blue!70] (0,0)--(0.35,0.95) node[left=1pt]{$\tau$};
    \draw[->,very thick,black] (0,0)--(1.35,0.95) node[right=1pt]{$\tau+1$};
    \draw[->,black,densely dashed] (0.35,0.95)--(1.30,0.95);
    \node[black] at (0.85,1.13) {\scriptsize $+1$};
    \node at (0.55,-0.6) {$T:\ \tau\to\tau+1$};
  \end{scope}
  % ----- S : tau -> -1/tau  (swap the two generators) -----
  \begin{scope}[shift={(3.6,0)}]
    \lat
    \draw[->,thick,black!55] (0,0)--(1,0) node[below right=-1pt]{$1$};
    \draw[->,very thick,blue!70] (0,0)--(0.35,0.95) node[left=1pt]{$\tau$};
    \draw[<->,thick,black] (0.80,0.18) to[bend left=45] (0.47,0.80);
    \node[black] at (1.02,0.66) {\scriptsize swap};
    \node at (0.55,-0.6) {$S:\ \tau\to-1/\tau$};
  \end{scope}
\end{tikzpicture}
\caption{The generators $T$ and $S$ acting on the lattice basis $\{1,\tau\}$, the two periods of the torus. $T:\tau\to\tau+1$ is a shift on the lattice by one period, while $S:\tau\to-1/\tau$ swaps the two generators, up to a rescaling that restores unit length.}
\label{fig:st-cycles}
\end{figure}

These two operations generate the \SLZ{} group, acting as $\tau \to \frac{a\tau+b}{c\tau+d}$ with $ad-bc=1$ (Figure~\ref{fig:st-cycles}).
The torus arises from periodic identifications on $S^1 \times S^1$, which fix the underlying lattice but not a choice of basis. 
The \SLZ{} symmetry is what relates the equivalent bases, whereas each basis consists of two fundamental 1-cycles of $T^2$.
A general 1-cycle is an integer combination of these, labelled by a pair $(p,q)$, which may be understood as winding $p$ times around one circle and $q$ times around the other.
When $p$ and $q$ are coprime it is considered a primitive $(p,q)$ cycle. 
Otherwise, writing $d \equiv \gcd(p,q)$, it is $d$ disjoint copies of the primitive $(p/d, q/d)$ cycle.
\SLZ{} permutes these primitive 1-cycles among themselves.

\begin{figure}[ht]
\centering
\includegraphics[width=0.25\textwidth]{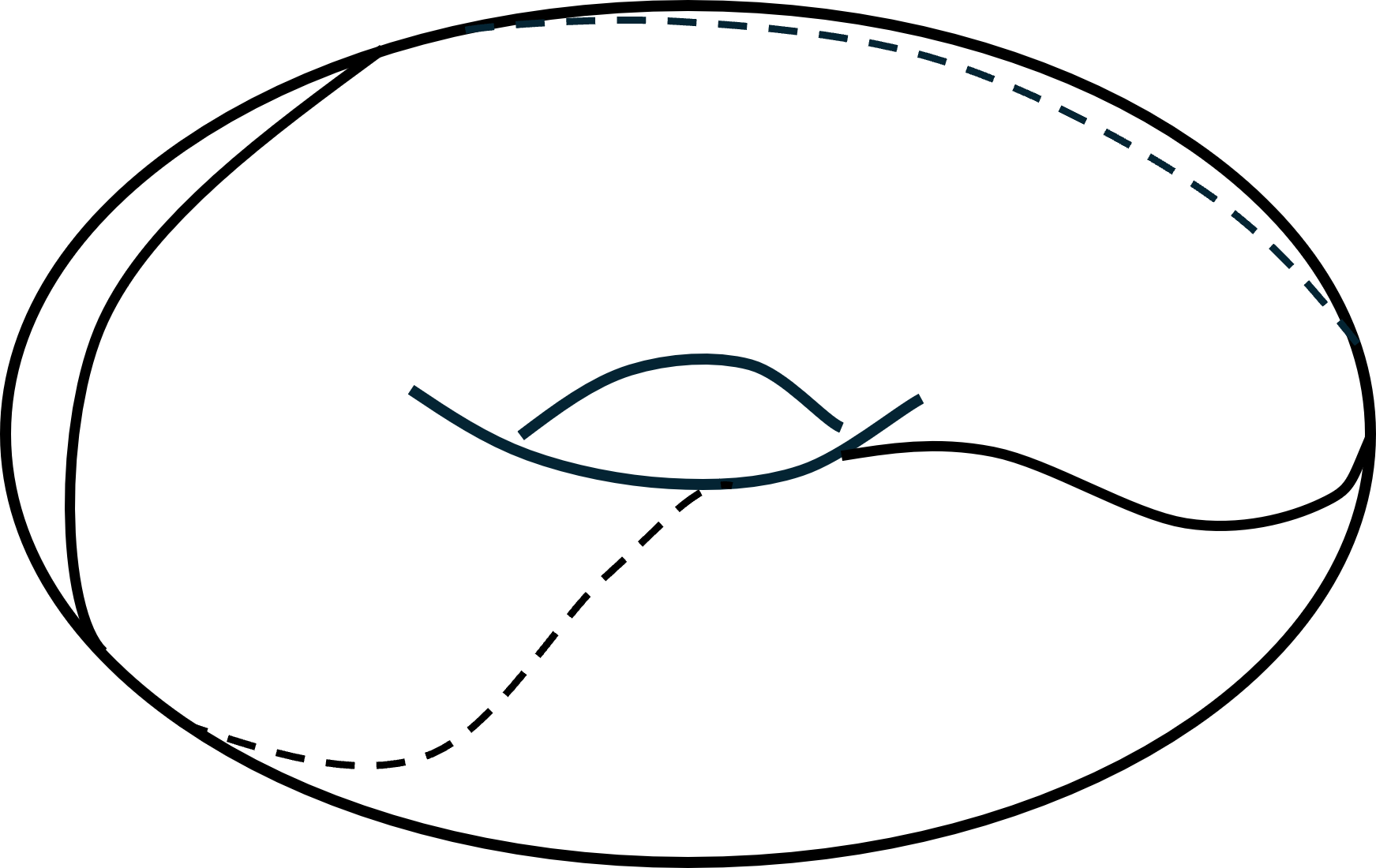}
\caption{A $(p,q)$ cycle on the torus that carries the 12d reduction. A $(p,q)$ 7-brane can be understood as the compactification of the 12d Kaluza-Klein monopole on $T^2$ whose Taub-NUT circle is wrapped on such a $(p,q)$ cycle.
Here being displayed is a $(1,1)$ cycle.}
\label{fig:pq-cycle}
\end{figure}

We can now draw the two threads together. 
We have seen that the type IIB D7 brane is obtained by reducing the 12d KK-monopole on $T^2$, where the Taub-NUT circle is identified with one of the $S^1$ factors of $T^2$, which we can take to be the $(1,0)$ cycle.
Furthermore, we identified the type IIB $S, T$ generators of \SLZ{} with large diffeomorphisms of the $T^2$.
Acting on the torus, these are precisely the relabellings of its 1-cycles introduced above: they carry the $(1,0)$ cycle to a general $(p,q)$ cycle, and with it the D7 brane to a $(p,q)$ 7-brane.
It is therefore natural to identify the $(p,q)$ 7-brane as the 12d KK-monopole whose Taub-NUT circle is wrapped on the $(p,q)$ cycle of $T^2$ (Figure~\ref{fig:pq-cycle}).

\subsection{D(-1) and D0 Interpreted as 12d and 11d pp-Waves}
The other $\frac{1}{2}$-BPS solution of the axio-dilaton action \eqref{IIB axio-dilaton action} is the D(-1) instanton, which has been recognised as a 12d pp-wave \cite{Tseytlin:1996ne}.
We now discuss this in the context of the known relation between the D0 brane and an 11d pp-wave.

Pp-waves are solutions of the vacuum Einstein equations following from the Einstein-Hilbert action, describing a gravitational wave propagating at the speed of light.
In $d$ dimensions, written in light-cone coordinates $v,u = z\pm t$, they take the form
\begin{equation}
ds^2 = dudv+(H-1)du^2+\sum_{i=1}^{d-2}dx_i^2,\quad
\nabla^2_{d-2} H(\vec{x})=0,
\label{d-dim pp-wave}
\end{equation}
where the wave profile $H$ is a harmonic function of the $d-2$ transverse coordinates $\vec x$, and the geometry is asymptotically flat as $H\to1$.
For a point source in the transverse space it is standard to take
\begin{equation}
H=1+\frac{Q}{r^{d-4}},
\end{equation}
with $r=|\vec x|$ the transverse radial coordinate.
The constant $Q$ sets the strength of the wave: it is the amplitude of the only metric component deviating from flat space, $g_{uu}=H-1=Q/r^{d-4}$, so $Q\to0$ recovers Minkowski space.
For our purposes, it is in fact more convenient to write \eqref{d-dim pp-wave} instead with explicit $t, z$ coordinates, as
\begin{equation}
ds^2 = - H^{-1}dt^2 + H[dz + (H^{-1} -1)dt]^2 + \sum_{i=1}^{d-2} dx_i^2.
\label{d-dim pp-wave-t-z}
\end{equation}

\paragraph{D0 brane from reduction of the 11d pp-wave}
By specialising \eqref{d-dim pp-wave-t-z} to 11d, we have the 11d pp-wave solution
\begin{equation}
ds_{11}^2
=
-H^{-1}dt^2+H[dz+(H^{-1}-1)dt]^2
+\sum_{i=1}^9 dx_i^2,\quad
H=1+\frac{Q}{r^7}.
\end{equation}
This can be reduced to 10d by comparison with the KK reduction ansatz \eqref{string frame KK reduction ansatz}.
Doing so we find precisely the 10d D0 brane solution in string frame \cite{Stelle:1998xg}
\begin{equation}
ds_{10,string}^2
=-H^{-1/2} dt^2+H^{1/2}ds_9^2,\quad
e^{\Phi} = H^{3/4},\quad
(C_1)_0 = H^{-1}-1.
\end{equation}
Like the relation between the 11d KK-monopole and 10d D6 brane in the type IIA theory, the connection between the 11d pp-wave and the D0 brane is part of the established S-duality between the type IIA theory and M-theory. Despite the absence of a 12d supergravity theory, this story also has a similar analogue on the type IIB side, between a 12d pp-wave and the type IIB D(-1) solution.

\paragraph{D(-1) instanton from reduction of the 12d pp-wave}
We now discuss the parallel story on the type IIB side. 
Specialising \eqref{d-dim pp-wave-t-z} to 12d, we find
\begin{equation}
ds_{12}^2 = - H^{-1}dt^2 + H[dz_1 + (H^{-1}-1)dt]^2 + \sum_{i=1}^{10} dx_i^2.
\end{equation}
Performing a Wick rotation via $t \equiv iz_2$ yields
\begin{equation}
ds_{12}^2
= \sum_{i=1}^{10} dx_i^2 + H^{-1}dz_2^2 + H[dz_1 + i(H^{-1}-1)dz_2]^2,
\end{equation}
which, in coordinates $(x^m, z_1, z_2)$, corresponds to the 12d metric embedding
\begin{equation}
\widehat{g}_{MN} = 
\begin{pmatrix}
\delta_{mn} & 0 \\
0 & \widehat{g}_{ab}
\end{pmatrix},\quad
\widehat{g}_{ab} =
e^{\Phi}\begin{pmatrix}
1 & C_0 \\
C_0 & C_0^2 + e^{-2\Phi}
\end{pmatrix},
\end{equation}
where
\begin{equation}
e^\Phi = H, \quad
C_0 = i (H^{-1}-1).
\end{equation}
This precisely reduces to the D(-1) solution of the type IIB axio-dilaton sector.
Here the axion $C_0$ is purely imaginary, which is a consequence of the Wick rotation we performed to keep the metric real.
It is also customary to define a real scalar $\mathcal C \equiv -i C_0$, which is the axion entering the Euclidean type IIB theory with a wrong sign kinetic term \cite{Gibbons:1995vg}.

One may view the Euclidean D(-1) as a ``slice'' of the 10d homogeneous wavefront of a 12d pp-wave.
The momentum of the wave is the D(-1) charge.
Recent investigations of the IKKT matrix model \cite{Hartnoll:2024csr,Komatsu:2024bop} reveal a type IIB supergravity background with axio-dilaton and the 3-form turned on, which admits a similar 12d interpretation.\footnote{
The mass-deformed model is dual to a probe D1 brane in the Einstein-frame-flat background $ds_{10}^2 = \sum_i dx_i^2$, $e^\Phi = 1 - \frac{\mu^2}{32}\left(\sum_{A=1}^7 x_A^2 + 3\sum_{a=9}^{10}x_a^2\right)$, $H_3 = \mu dx^8\wedge dx^9\wedge dx^{10}$, valid in the ellipsoidal region where $e^\Phi\ge0$ \cite{Hartnoll:2024csr}. Uplifting to 12d (Wick rotating to keep $ds^2$ real) gives the Brinkmann-form metric $ds_{12}^2 = 2dudv + e^\Phi du^2 + \sum_{i=1}^{10}dx_i^2$, whose only nonvanishing Ricci component is $R_{uu}=\mu^2/2$. Since $e^\Phi$ is not harmonic it is not a pp-wave, but it solves 12d gravity coupled to a 4-form, $S = \int d^{12}x\sqrt{-\widehat{g}}\left(\widehat{R}-\tfrac12|\widehat{F}_4|^2\right)$ with $\widehat{F}_4 = \mu du\wedge dx^8\wedge dx^9\wedge dx^{10} = du\wedge H_3$.
}

\subsection{Supersymmetry}
We now discuss a further consistency check for the relation between 12d gravity and the type IIB axio-dilaton sector, namely how the $\tfrac{1}{2}$ BPS condition of the latter can be obtained by reducing the covariantly-constant-spinor equation of the former.

Disregarding the type IIB supergravity action for the moment, let us consider a generic theory in $d$ dimensions.
The action with kinetic terms for the metric, form fields, and the gravitino takes the form
\begin{equation}
S = \int d^d x  \sqrt{-g}\left[
R
- \frac{1}{2}\sum_n \frac{1}{n!} |F_n|^2
- i \bar{\psi}_M \Gamma^{MNP} \nabla_N \psi_P
\right] + \dots.
\label{generic sugra action}
\end{equation}
Under a local supersymmetry transformation parameterised by a spinor $\epsilon$, one can write down a general equation for the variation of the gravitino from covariance alone
\begin{equation}
\delta\psi_M
= \left\{
\nabla_M + \sum_n \left[
c_n (F_{n})_M{}^{N_1 N_2 \cdots} \Gamma_{N_1 N_2 \cdots}
+ d_n (F_{n})_{N_1 N_2 \cdots} \Gamma_M{}^{N_1 N_2 \cdots}
\right]
\right\}\epsilon,
\label{generic gravitino variation}
\end{equation}
with summation over the form fields of the given theory, and $c_n, d_n$ are coefficients to be determined.
The Killing spinor equation would then be given by $\delta\psi_M=0$.

Upon dimensional reduction, the higher-dimensional fields decompose into those of the lower-dimensional theory, and reducing the gravitino variation \eqref{generic gravitino variation} reproduces its Killing spinor equation.
In our case the higher-dimensional theory is pure gravity, with no form fields, so its gravitino variation \eqref{generic gravitino variation} is simply $\nabla_M\epsilon$.
The full lower-dimensional Killing spinor equation, flux and connection terms included, should then be reproduced by reducing this covariant derivative alone.

\paragraph{From 11d to type IIA.}
Let us consider the 11d covariant constancy equation $\widehat{\nabla}_M \epsilon = 0$.
Splitting the 11d indices into $x^M = (x^m, x^{11})$, we have
\begin{equation}
\begin{aligned}
\widehat{\nabla}_m \epsilon 
&= \nabla_m \epsilon + \frac{1}{4} \widehat{w}_{mBC} \Gamma^{BC} \epsilon = 0,\\  
\widehat{\nabla}_{11} \epsilon 
&= \frac{1}{4} \widehat{w}_{11,BC} \Gamma^{BC} \epsilon = 0.
\end{aligned}
\end{equation}
Substituting the reduced spin connection (see \cite{Ortin:2015hya}) on the standard Kaluza-Klein ansatz \eqref{string frame KK reduction ansatz}, they become
\begin{equation}
\begin{aligned}
\widehat{\nabla}_m \epsilon
&= e^{\Phi/3}\left[ \nabla_m - \frac{1}{6} \Gamma_m{}^n \partial_n \Phi + \frac{1}{4} e^\Phi F_{mn} \Gamma^n \Gamma^{11} \right] \epsilon = 0,\\
\widehat{\nabla}_{11} \epsilon
&=\frac{1}{4} e^{\Phi/3} \left[ -\frac{1}{2} e^{\Phi} F_{mn} \Gamma^{mn} - \frac{4}{3}\partial_m \Phi \Gamma^m \Gamma^{11} \right] \epsilon = 0.
\end{aligned}
\end{equation}
We can substitute the $\widehat{\nabla}_{11} \epsilon = 0$ equation into the $\widehat{\nabla}_m \epsilon = 0$ equation to find
\begin{equation}
\nabla_m \epsilon = e^{\Phi/3} \left[ \nabla_m + e^\Phi \left(\frac{1}{8}F_{mn}\Gamma^n - \frac{1}{16}F_{nr}\Gamma_m{}^{nr}\right) \Gamma^{11} + \frac{1}{6}\partial_m\Phi \right] \epsilon = 0.
\end{equation}
The flux term can be simplified:
\begin{equation}
\begin{aligned}
\frac{1}{8}F_{mn}\Gamma^n - \frac{1}{16}F_{nr}\Gamma_m{}^{nr}
&= -\frac{1}{16}F_{nr}(\Gamma_m{}^{nr} - 2\delta_m^{[n}\Gamma^{r]})\\
&= - \frac{1}{8}F_{nr}\{\Gamma_m,\Gamma^{nr}\}\\
&= - \frac{1}{8} F_{nr} \Gamma_m{}^{nr}.
\end{aligned}
\end{equation}
Then performing a redefinition $\epsilon \to e^{-\Phi/6} \epsilon$ to remove the $\partial\Phi$ term, we arrive at
\begin{equation}
\left(
\nabla_m - \frac{1}{8}e^\Phi F_{nr} \Gamma_m{}^{nr}\Gamma^{11}
\right)\epsilon = 0,
\end{equation}
which is precisely the type IIA Killing spinor equation in the gravity, KK-vector and dilaton sector \cite{Becker:2006dvp}.
Hence the 10d $\tfrac{1}{2}$ BPS condition may be understood from 11d covariance.
Now we present an analogous story between 12d covariance and the type IIB theory.

\paragraph{From 12d to type IIB.}
A 12d covariant derivative $\widehat{\nabla}^{(12)}_M\epsilon = 0$ restricted to 10d coordinates $x^m$ takes the form
\begin{equation}
\nabla^{(12)}_m\epsilon
=\left(
\nabla^{(10)}_m
+\frac{1}{2}\widehat{w}_m{}^{11,n}\Gamma_{11,n}
+\frac{1}{2}\widehat{w}_m{}^{12,n}\Gamma_{12,n}
+\frac{1}{2}\widehat{w}_m{}^{11,12}\Gamma_{11,12}\right)\epsilon = 0,
\label{12d covariant derivative}
\end{equation}
Using the 12d metric ansatz, we find the spin connections
\begin{equation}
\widehat{w}_m{}^{11,n}=\widehat{w}_m{}^{12,n}=0,\quad
\widehat{w}_m{}^{11,12} =Q_m.
\end{equation}
The 12d covariant derivative \eqref{12d covariant derivative} thus becomes
\begin{equation}
\left[
\nabla_m + \frac{1}{2} Q_m \Gamma_{11,12}
\right]\epsilon.
\end{equation}
Note that on a Euclidean torus
\begin{equation}
\Gamma_{11,12} = \Gamma_{11}\Gamma_{12},\quad
(\Gamma_{11,12})^2 =-\Gamma_{11}^2\Gamma_{12}^2 = -1,
\end{equation}
so $\Gamma_{11,12}$ is, up to a similarity transformation, $i$ times the U(1) generator.
After performing a similarity transformation one obtains the axio-dilaton sector Killing spinor equation \eqref{9+1 IIB axio-dilaton Killing spinor equations}.
This would also hold had we instead compactified on a torus of signature (1,1), to obtain a Euclidean type IIB theory.
In that case, a factor of $i$ arises both in identifying the U(1) generator from $\Gamma_{11,12}$ and in defining the Euclidean compact scalar $C_0 = i\mathcal C$, so that the 12d covariant derivative again reproduces the type IIB gravitino variation.

\section{Does the Story Continue? A Covariant Unification in 12d}
\label{sec:covariant-unification}

In the previous section we saw that the D(-1) and D7 solutions lift to 12d, and that, if a 12d theory existed, reducing its covariantly constant spinor condition on a torus would reproduce the type IIB Killing spinor equations.
It is natural to ask whether other brane-like solutions of type IIB lift in the same way.

\subsection{Brane-Like Solutions}
\label{sec:general brane solns}
Working in the Einstein frame, with $F_n$ an n-form and the action
\begin{equation}
S = \int d^Dx\sqrt{-g}\left[
    R - \frac{1}{2}(\partial\phi)^2 - \frac{1}{2}e^{\alpha\phi}|F_n|^2
\right],
\end{equation}
the equations of motion, together with the Bianchi identity (assuming $F = dA$), are given by \cite{Stelle:1998xg}
\begin{equation}
\begin{aligned}
R_{MN}
&= \frac{1}{2}\partial_M\phi\partial_N\phi + e^{\alpha\phi}S_{MN}\big|_F,\\
dF&=0,\\
d[e^{\alpha\phi}*_D F] &=0,\\
\Box\phi&=\frac{\alpha}{2}e^{\alpha\phi}|F|^2,
\end{aligned}
\end{equation}
where
\begin{equation}
S_{MN}\big|_{F}\equiv
\frac{1}{2}\left[
    \frac{1}{(n-1)!}F_{M...}F_N^{...}
    -\frac{n-1}{D-2}\left(\frac{1}{n!}F^2\right)g_{MN}
\right],
\end{equation}
and $\Delta\equiv \alpha^2 +\frac{2 d \tilde{d}}{D-2}$, with $\tilde{d} \equiv D-d-2$.
There exists the electric solution, defined as the case where the form-field potential has legs along the worldvolume directions,
\begin{equation}
\begin{aligned}
ds_D^2
&=H^{-\frac{4\tilde d}{\Delta(D-2)}}ds_{d}^2+
H^{\frac{4d}{\Delta(D-2)}}ds_{D-d}^2,\quad
d = n-1,\\
F_{\mu_0\mu_1...\mu_{n-2},m}
&=\frac{2}{\sqrt\Delta}\varepsilon_{\mu_0\mu_1...\mu_{n-2}}\partial_m (H^{-1}),\\
e^\phi
&=H^{2\alpha/\Delta},
\end{aligned}
\end{equation}
as well as the magnetic solution, defined as the case where the field strength has its legs along the transverse directions,
\begin{equation}
\begin{aligned}
ds_D^2
&=H^{-\frac{4\tilde d}{\Delta(D-2)}}ds_{d}^2+
H^{\frac{4d}{\Delta(D-2)}}ds_{D-d}^2,\quad
d = D-n-1,\\
F_{m_1...m_n}
&= \frac{2}{\sqrt{\Delta}}\varepsilon_{m_1...m_n m}\partial_m H,\\
e^\phi &= H^{-2\alpha/\Delta}.
\end{aligned}
\end{equation}
In 11d supergravity, type IIA, and type IIB, one finds $\Delta = 4$ for all branes.
We provide a catalog of the various brane solutions in Appendix~\ref{sec:catalog-of-branes}.

One can readily see that there do not exist straightforward ways to embed the type IIB metric, scalars, and form fields into a 12d metric and form fields.
At the level of supergravity, 12d fields transform under the little group SO(10) while the 10d fields transform under SO(8).
In Table~\ref{tab:10d-12d-dof} we summarise the 10d massless on-shell degrees of freedom in type IIB, alongside the on-shell degrees of freedom of the corresponding 12d field of the same form rank.

\begin{table}[h]
\centering
\begin{tabular}{l|ccccccc}
\hline
 & metric & scalar & 1-form & 2-form & 3-form & 4-form & 5-form \\
\hline
Type IIB $SO(8)$ & $1\times 35$ & $2\times 1$ & $0\times 8$ & $2\times 28$ & $0\times 56$ & $1\times 35$ (self-dual) & $-$ \\
12d $SO(10)$ & $54$ & $1$ & $10$ & $45$ & $120$ & $210$ & $126$ (self-dual) \\
\hline
\end{tabular}
\caption{
On-shell bosonic degrees of freedom of the type IIB fields, given as (number of fields) $\times$ (dimension of the $SO(8)$ little-group representation), compared with the dimension of the $SO(10)$ representation of a 12d field of the same form rank.
}
\label{tab:10d-12d-dof}
\end{table}
As the table shows, the type IIB degrees of freedom do not fill out the SO(10) representations of the 12d fields. 
Type IIB therefore cannot arise as the dimensional reduction of a genuine 12d (super)gravity that realises the full 12d spacetime symmetries.

This does not, however, exhaust the 12d interpretation.
Already in the axio-dilaton sector the embedding into the 12d metric was only partial: the two off-diagonal Kaluza-Klein 1-forms were set to zero, so the full 12d metric was never reconstructed. It nonetheless provided a 12d perspective on the D(-1) and D7 branes, as we saw in the previous section.
In the same spirit, the following subsection organises the type IIB fields into 12d objects \emph{without} realising the full 12d spacetime symmetries: not all degrees of freedom of the 12d graviton and form fields are excited, so the construction is best understood as a covariant repackaging of the 10d theory rather than the reduction of a genuine 12d one.

\subsection{A Covariant Unification in 12d}
\label{sec:a-covariant-unification-in-12d}

The various brane-related evidence for an effective 12d interpretation of the type IIB theory suggests a very specific 12d interpretation of the axio-dilaton sector.
In addition, one seeks a 12d interpretation of the type IIB RR and NSNS form fields.
In this section, we gather the various 12d insights obtained in the previous section from the type IIB branes, and propose a covariant 12d unification of the $\SLR{}\times SO(9,1)$ form fields.
We will use hats $\widehat{\phantom{x}}$ to denote fields in the higher dimension.
The 10d metric $g_{mn}$ with $m,n=0,1,...,9$ will be interpreted as embedded inside a 12d metric $\widehat{g}_{MN}$, with $M,N=0,1,...,11$. We will use $\widehat R$ to denote the Ricci scalar computed with $\widehat{g}_{MN}$, and use $\widehat{F}_{n+1} = d\widehat{C}_{n}$ to denote form fields in 12d.

Let the 12d coordinates be parameterised by $(x^m,z_1,z_2)$, and let $\widehat{g}_{ab}$ be the $2\times 2$ metric on the torus.\footnote{
We orient the torus so that $dz_2\wedge dz_1$ is the positively-oriented volume element, i.e. $\int_{T_2}dz_1\wedge dz_2 = -\mathrm{Vol}(T_2)$. This fixes the sign of the 12d Hodge dual $*_{12}$, and hence the signs in the self-duality relation \eqref{12d interpretation of 10d self-duality} and in the Chern-Simons term.
}
The key insight from the previous section is that the axio-dilaton sector of type IIB may arise from the 12d metric embedding given by
\begin{equation}
\widehat{g}_{MN}=\begin{pmatrix}
g_{mn} & 0 \\
0 &\widehat{g}_{ab}
\end{pmatrix},\quad
\widehat{g}_{ab}
=\frac{1}{\tau_2}\begin{pmatrix}
1& \tau_1\\
\tau_1 & \tau_1^2+\tau_2^2
\end{pmatrix}.
\label{12d metric ansatz}
\end{equation}
In particular,
\begin{equation}
\begin{aligned}
\frac{1}{2\kappa^2}\int d^{10}x\sqrt{-g}\left(
R - \frac{|\nabla \tau|^2}{2\tau_2^2}
\right)
&=
\frac{\mathrm{Vol}(T_2)}{2\kappa_{12}^2}\int d^{10}x \sqrt{-\widehat{g}} \widehat{R}\\
&=\frac{1}{2\kappa_{12}^2}\int_{T_2}dz_1 dz_2\int d^{10}x\sqrt{-\widehat{g}}\widehat{R},
\end{aligned}
\label{axio-dilaton 12d uplift}
\end{equation}
where we have schematically defined $\kappa_{12}$ by 
\begin{equation}
\frac{1}{\kappa^2} = \frac{\mathrm{Vol}(T_2)}{\kappa_{12}^2}
=\frac{1}{\kappa_{12}^2}\int_{T_2}*_2 1,\quad
[\kappa_{12}^2] = L^{10}.
\end{equation}
The 3-form field strengths form an \SLR{} doublet.
To unify them in 12d we define a 12d 4-form field strength with exactly one leg on the torus:
\begin{equation}
\widehat{F}_4=d\widehat C_3,\quad
\widehat C_3 \equiv B_2\wedge dz_1 + C_2\wedge dz_2.
\end{equation}
Using the 12d metric ansatz \eqref{12d metric ansatz}, the 10d 3-form field strengths and their axio-dilaton couplings follow from contracting $\widehat F_4$:
\begin{equation}
\begin{aligned}
|\widehat{F}_4|^2\bigg|_{\widehat{g}_{MN}}
&=\widehat{g}^{z_1 z_1}|H_3|^2+2\widehat{g}^{z_1 z_2}(F_3\cdot H_3)+\widehat{g}^{z_2 z_2}|F_3|^2\\
&=\left(
e^{-\Phi}|H_3|^2 + e^\Phi|F_3-C_0H_3|^2
\right)\bigg|_{g_{mn}}.
\end{aligned}
\end{equation}
Finding a 12d interpretation for the 5-form sector is trickier. 
One observes that in 10d, the composite \SLR{} singlet 5-form 
\begin{equation}
\tilde F_5 = F_5 - \frac12 C_2 \wedge H_3 + \frac{1}{2}B_2\wedge F_3    
\end{equation}
cannot be sensibly constructed in 12d, due to the lack of a pair of form field potential and strength with ranks that sum to 5 in 12d.
However, the 10d five-form field strength admits two possible uplifts to 12d:
\begin{equation}
\widehat{F}_5 = F_5,\quad \widehat{F}_7=F_5\wedge dz_1 \wedge dz_2.
\end{equation}
Then note that
\begin{equation}
\begin{aligned}
|\widehat{F}_5|^2\bigg|_{\widehat{g}_{MN}} 
&= |F_5|^2\bigg|_{g_{mn}},\\
|\widehat{F}_7|^2\bigg|_{\widehat{g}_{MN}}
&=\det(\widehat{g}^{ab})|F_5|^2\bigg|_{g_{mn}} = |F_5|^2\bigg|_{g_{mn}}.
\end{aligned}
\end{equation}
Then we may define a 12d 7-form
\begin{equation}
\widetilde{\widehat{F}}_7 \equiv \widehat{F}_7 - \frac{1}{2}\widehat{C}_3 \wedge \widehat{F}_4
\end{equation}
that exactly supplies the type IIB composite 5-form contribution upon contraction in 12d:
\begin{equation}
|\widetilde{\widehat{F}}_7|^2\bigg|_{\widehat{g}_{MN}} =
\det(\widehat{g}^{ab})|\tilde F_5|^2\bigg|_{g_{mn}}.
\end{equation}
The additional utility of the composite seven-form is that it lets us write the genuine 10d self-duality condition on the composite five-form $\tilde F_5$ as a 12d Hodge duality condition
\begin{equation}
\widetilde{\widehat{F}}_7 = -*_{12}\tilde F_5
\quad
\Leftrightarrow\quad
\tilde F_5 = *_{10}\tilde F_5.
\label{12d interpretation of 10d self-duality}
\end{equation}
The 10d Chern-Simons term can also be obtained using the 12d potentials and their corresponding field strengths:
\begin{equation}
\frac{1}{2}\int_{T_2\times \mathbb{R}^{1,9}} \widehat{C}_4\wedge \widehat{F}_4\wedge\widehat{F}_4=
\mathrm{Vol}(T_2)\int_{\mathbb{R}^{1,9}}C_4\wedge H_3\wedge F_3.
\end{equation}
We note that reproducing the type IIB Chern-Simons term in 12d necessitates both the 4-form field strength $\widehat F_4$ and the 4-form potential $\widehat C_4$, whose field strength is the 5-form $\widehat F_5$.

One can write down a ``12d''\footnote{
Not dynamical 12d, but dynamical 10d times a 2-torus.
}
action
\begin{equation}
\begin{aligned}
S_{IIB}=
S_{\text{``12''}}
= \frac{1}{2\kappa_{12}^2}\int_{T_2}dz_1 dz_2\int d^{10}x
\sqrt{-\widehat{g}}
\left(
\widehat{R} - \frac{1}{2}|\widehat F_4|^2-\frac{1}{4}|\widetilde{\widehat{F}}_7|^2
\right)-\frac{1}{8\kappa_{12}^2}\int_{T_2\times \mathbb{R}^{1,9}}\widehat{C}_4\wedge \widehat{F}_4\wedge \widehat{F}_4.
\end{aligned}
\label{12d action}
\end{equation}
The composite seven-form $\widetilde{\widehat{F}}_7$ does not arise from an independent degree of freedom: it is fixed by the self-duality relation \eqref{12d interpretation of 10d self-duality}, $\widetilde{\widehat{F}}_7 = -*_{12}\tilde F_5$. 
The action \eqref{12d action} is exactly the Einstein frame type IIB action \eqref{eq:sl2r_invariant_einstein_frame_action}.
The 2d integrand is just a repackaging of $1/\kappa^2$, and it is likely that $\widehat C_3, \widehat C_4$ do not furnish independent degrees of freedom, as has been discussed in \cite{Tseytlin:1996ne}.

\subsection{Uplift of Type IIB Branes}
Using the effective action \eqref{12d action}, we now uplift the type IIB brane solutions to 12d.\footnote{The brane solutions are listed in Appendix~\ref{sec:catalog-of-branes}.}
We must note that these uplifts do not solve the full 12d Einstein equation.
This is a manifestation of the fact that the equation of motion obtained in 12d by varying the volume of the torus is not imposed in 10d, where it's not a dynamical field.
We accordingly treat the uplifts as a 12d repackaging of the 10d backgrounds, rather than as genuine 12d solutions.

\paragraph{F1}
\begin{equation}
\begin{aligned}
ds^2
&= H^{-3/4}ds_{1,1}^2
+ H^{1/4}ds_8^2
+H^{-1/2}dz_1^2
+H^{1/2}dz_2^2,\\
(\widehat{F}_4)_{\mu_0 \mu_1 m z_1}
&= \varepsilon_{\mu_0 \mu_1}\partial_m H^{-1}.
\end{aligned}
\end{equation}

\paragraph{D1}
\begin{equation}
\begin{aligned}
ds^2
&= H^{-3/4}ds_{1,1}^2
+ H^{1/4}ds_8^2
+H^{1/2}dz_1^2
+H^{-1/2}dz_2^2,\\
(\widehat{F}_4)_{\mu_0 \mu_1 m z_2}
&= \varepsilon_{\mu_0 \mu_1}\partial_m H^{-1}.
\end{aligned}
\end{equation}

\paragraph{NS5}
\begin{equation}
\begin{aligned}
ds^2
&= H^{-1/4}ds_{1,5}^2
+ H^{3/4}ds_4^2
+H^{1/2}dz_1^2
+H^{-1/2}dz_2^2,\\
(\widehat{F}_4)_{m_1 m_2 m_3 z_1}
&= \varepsilon_{m_1 m_2 m_3 m}\partial_m H.
\end{aligned}
\end{equation}

\paragraph{D5}
\begin{equation}
\begin{aligned}
ds^2
&= H^{-1/4}ds_{1,5}^2
+ H^{3/4}ds_4^2
+H^{-1/2}dz_1^2
+H^{1/2}dz_2^2,\\
(\widehat{F}_4)_{m_1 m_2 m_3 z_2}
&= \varepsilon_{m_1 m_2 m_3 m}\partial_m H.
\end{aligned}
\end{equation}

\paragraph{Comments on role of dilaton and F-theory}
Usually, in supergravity theories, the dilaton is related to the volume of the compact manifold.
In type IIB, however, the compact torus has fixed volume: as the radius of one circle grows, the other shrinks.
The natural place for the non-compact scalar dilaton to live is the radius of one of the circles $R_1 \sim e^{\Phi}$, while the radius of the other circle scales inversely, $R_2 \sim e^{-\Phi}$.
This is apparent by examining the conformal factors on $dz_1^2$ and $dz_2^2$ in the 12d uplifts above.
Following this logic, type IIB backgrounds cannot have a genuine 12d decompactifying limit, because taking the decompactifying limit of one circle immediately compactifies the other circle.
The exception is the D(-1)-instanton uplift, where the pp-wave propagates inside a non-compact torus.
The KK-monopole uplift of the D7-brane is not an exception because a 12d KK-monopole is localised on $S^1 \times \mathbb{R}^3$: there is already a compact circle in 12d.

While the two circle radii scale inversely to each other, one may consider the special case where they are equal: $e^\Phi \propto e^{-\Phi}$.
This implies that the radii of both circles are finite constants hence the 10d dilaton is not excited, as in the case of the D3 background.
\section{The D3 Brane and Self-Duality}
\label{sec:d3-self-duality}
The low-energy dynamics of a D3 brane in a given type IIB supergravity background are described by the Dirac-Born-Infeld action supplemented by a Wess-Zumino term, with background metric, NSNS and RR fields entering through the appropriate pullbacks to the brane worldvolume \cite{Leigh:1989jq,Polchinski:1998rr}.
In particular, the bosonic sector of the D3 low-energy action takes the form \cite{Tseytlin:1996it}
\begin{equation}
\begin{aligned}
S_{D3}
&=\int d^4x \sqrt{-\det\big({g}_{mn} + e^{-\Phi/2} (\mathcal{F}_2)_{mn}\big)}\\
&\quad+\int \Big(C_4 + C_2\wedge \mathcal{F}_2 + \tfrac{1}{2} C_0 \mathcal{F}_2\wedge\mathcal{F}_2\Big)
+\text{higher order},\qquad
\mathcal{F}_2 \equiv dA_1+B_2 .
\end{aligned}
\label{D3-action}
\end{equation}
Throughout, the bulk supergravity fields are understood to be pulled back onto the brane worldvolume (in \eqref{D3-action} we pullback the Einstein frame bulk metric to the D3 worldvolume), where $m,n$ label worldvolume indices, $(\mathcal{F}_2)_{mn}$ are the components of the worldvolume 2-form $\mathcal{F}_2$, and $C_0$ is the RR axion.
The vector $A_m$ is the dynamical gauge potential on the brane worldvolume.
It was shown \cite{Tseytlin:1996it} that the action is invariant under electromagnetic duality only if the bulk background simultaneously undergoes an \SLZ{} transformation.
We derive this explicitly below.

To quadratic order in $\mathcal F_2$ the gauge sector of \eqref{D3-action} reads, in the notation of differential forms,
\begin{equation}
\int\left[\tfrac{1}{2}e^{-\Phi}\mathcal F_2\wedge\star\mathcal F_2+\tfrac{1}{2}C_0\mathcal F_2\wedge\mathcal F_2+C_2\wedge\mathcal F_2\right],\qquad \mathcal F_2 = dA_1+B_2.
\end{equation}
To dualise, we promote $\mathcal F_2$ to an independent field and add a Lagrange multiplier 
\begin{equation}
\tilde{F}_2 \wedge (\mathcal{F}_2 - B_2)
\end{equation}
to enforce $d(\mathcal F_2-B_2) = 0$, where $\tilde F_2 = d\tilde A_1$ is the field strength of the dual potential $\tilde A_1$.
Varying the now-independent $\mathcal F_2$ instead gives the algebraic relation
\begin{equation}
(e^{-\Phi}\star + C_0)\mathcal F_2 = -(\tilde F_2+C_2).
\label{D3-integrating-out-mathcal-F}
\end{equation}
We now integrate $\mathcal F_2$ out and ask whether the resultant D3 action is in the form \eqref{D3-action}, with $d\tilde A_1$ in place of $dA_1$. 
Using $\star^2=-1$, we have
\begin{equation}
(C_0+e^{-\Phi}\star)^{-1} = \frac{C_0-e^{-\Phi}\star}{C_0^2+e^{-2\Phi}},
\end{equation}
so \eqref{D3-integrating-out-mathcal-F} solves to $\mathcal F_2 = -\dfrac{C_0-e^{-\Phi}\star}{C_0^2+e^{-2\Phi}}(\tilde F_2+C_2)$. Substituting back, the gauge sector becomes the same Maxwell form for the dual field strength $\tilde F_2+C_2$, with the coupling inverted to $\tau'\equiv C_0'+ie^{-\Phi'}=-1/\tau$:
\begin{equation}
\begin{aligned}
\int\bigg[&\tfrac{1}{2}e^{-\Phi'}(\tilde F_2+C_2)\wedge\star(\tilde F_2+C_2)\\
&+\tfrac{1}{2}C_0'(\tilde F_2+C_2)\wedge(\tilde F_2+C_2)-\tilde F_2\wedge B_2\bigg].
\end{aligned}
\end{equation}
We read off that this is in the same form as the D3 action \eqref{D3-action}, with $\tilde A_1$ in place of $A_1$, provided we make the following field redefinitions
\begin{equation}
\tau\to-\frac{1}{\tau},\qquad
B_2 \to C_2,\qquad C_2\to -B_2.
\end{equation}
This is precisely the $S$ generator of the bulk \SLZ{} duality, realised on the worldvolume as electromagnetic duality.
The remaining generator, the axion shift $C_0\to C_0+1$, leaves the D3 action invariant and corresponds to $\theta\to \theta + 2\pi$ for the complexified coupling $\tau_{YM}\equiv \frac{\theta}{2\pi} + \frac{4\pi i}{g^2}$.
The full bulk \SLZ{} duality of type IIB is thus identified with the \SLZ{} duality of $\mathcal N=4$ SYM on the D3 worldvolume.

For comparison, we recall a similar story on the type IIA side.
The analogy is not direct because there is no analogue of the \SLZ{} duality in the type IIA theory.
The D2 action can be written as (here we pullback the string frame metric to the brane worldvolume) \cite{Schmidhuber:1996fy,Tseytlin:1996it}
\begin{equation}
S_{D2}
=\int d^3x \sqrt{-e^{-2\Phi}\det (G_{mn}+(\mathcal F_2)_{mn})}
+\int\left(C_3-C_1\wedge\mathcal F_2\right),\qquad
\mathcal F_2\equiv dA_1-B_2.
\end{equation}
We dualise the worldvolume vector into a scalar, paralleling the D3 treatment.
This is done by promoting $\mathcal F_2$ to an independent field and enforcing $d\mathcal F_2 = 0$ through a Lagrange multiplier, here a scalar $y$,
\begin{equation}
\int dy\wedge(\mathcal F_2 + B_2).
\end{equation}
Trading the independent $\mathcal F_{mn}$ for its dual vector through $\mathcal F_{mn}=\epsilon_{mnp}v^p$, so that $\tfrac12\mathcal F_{mn}\mathcal F^{mn}=-v_pv^p$, the action becomes%
\footnote{In three dimensions the Born-Infeld determinant collapses: for antisymmetric $\mathcal F_{mn}$ one has $\det(G_{mn}+\mathcal F_{mn})=\det G_{mn}(1+\tfrac12\mathcal F_{mn}\mathcal F^{mn})$, so the DBI term equals $e^{-\Phi}\sqrt{-G}\sqrt{1+\tfrac12\mathcal F_{mn}\mathcal F^{mn}}$.}
\begin{equation}
S_{D2}=\int d^3x\sqrt{-G}\left[e^{-\Phi}\sqrt{1-v_pv^p}-(C_1-dy)_pv^p\right]+\int\left(C_3 + B_2\wedge dy\right).
\end{equation}
Integrating out $v$ using its algebraic equation of motion
\begin{equation}
\frac{e^{-\Phi}v_p}{\sqrt{1-v_qv^q}}=-(C_1-dy)_p,
\label{D2-integrating-out-v}
\end{equation}
we find
\begin{equation}
S_{D2}=\int d^3x\sqrt{-G}\sqrt{e^{-2\Phi}+(C_1-dy)_p(C_1-dy)^p}+\int\left(C_3 + B_2\wedge dy\right).
\end{equation}
Finally, the identity $\det(aG_{mn}+bK_mK_n)=a^3\det G_{mn}(1+\tfrac{b}{a}K_pK^p)$ with $a=e^{-2\Phi/3}$, $b=e^{4\Phi/3}$ repackages the integrand into a single determinant
\begin{equation}
\begin{aligned}
S_{D2}
&=\int d^3x\sqrt{-\det\big(e^{-2\Phi/3}G_{mn} + e^{4\Phi/3}(C_1-dy)_m(C_1-dy)_n\big)}\\
&\quad+ \int\left(C_3 + B_2\wedge dy\right).
\end{aligned}
\end{equation}
As one interprets the worldvolume scalar $y$ as a 10d scalar pulled back onto the D2 worldvolume, this is the M2 action directly reduced on $S^1$ \cite{Bergshoeff:1987cm,Tseytlin:1996it,Schmidhuber:1996fy}. The bracketed metric is in the standard Kaluza-Klein form, and $C_3+B_2\wedge dy$ is the reduction of the 11d three-form.

Just as the D2 arises from the M2 reduced on a circle, if there were to exist a 12d effective framework, the D3 may arise from a 12d brane reduced on a two-torus $T^2$.
Suppose there exists a certain 5-brane in 12d, which would have its own worldvolume gauge theory.
Wrapping such a 5-brane on $T^2$ parametrised by $\tau$ would give a 3-brane in 10d that is an \SLZ{} singlet.
The worldvolume gauge theory of the resultant 3-brane would be related to the modulus $\tau$ of the torus, and \SLZ{} transformations on such $\tau$ would act both on the bulk and on the worldvolume, as we have seen in the D3 case.
Such a relationship is realised in the known example of the M5 brane in 11d, whose worldvolume theory is the 6d $(2,0)$ superconformal theory with a self-dual 2-form \cite{Witten:1995zh}.
The $\mathcal N = 4$ SYM on the D3, and hence the \SLZ{} duality is understood to be related to the 6d $(2,0)$ theory on $T^2$ \cite{Witten:2007ct}. 
The unsatisfactory aspect of this story is again that a 9d detour is required.
The same logic applies to a hypothetical 3-brane in 12d: the pullback of the 12d metric carries the torus modulus $\tau$ into its worldvolume action, so \SLZ{} again acts as a worldvolume transformation.

Whether such a connection exists in 12d remains an open question, and a closer look at the D3's degrees of freedom shows why it is not straightforward.
A naive count of degrees of freedom is encouraging: a 3-brane in 12d has eight transverse directions, and the D3 worldvolume theory carries eight bosonic degrees of freedom, six from its scalars and two from the vector field.
But the two vector degrees of freedom cannot be directly interpreted as transverse coordinates.
One can wrap the D3 on a circle so that its worldvolume vector decomposes into two scalars \cite{Jatkar:1996np,Kar:1997cx}, but this simply reproduces the standard relation between M-theory on $T^2$ and 9d supergravity \cite{Green:1997as}.

\section{Effective Actions and their 12d Consistencies}
\label{sec:effective-actions}
The type IIB supergravity \eqref{eq:sl2r_invariant_einstein_frame_action} is understood as the low-energy effective field theory of the type IIB string theory.
Under higher-derivative and stringy corrections, the \SLR{} symmetry is believed to be broken to the discrete group \SLZ{} \cite{Schwarz:1995dk}.
If the type IIB theory admits a 12d interpretation, that interpretation and its implications for effective actions must be consistent with the type IIB effective action and \SLZ{}.

In this section, we discuss the current understanding of \SLZ{} as arising from higher-derivative corrections, and their 12d interpretations.
We will begin by reviewing how one obtains higher-derivative corrections in supergravity from perturbative string theory, in particular how the \SLZ{} invariance arises non-perturbatively from the D(-1) backgrounds.
We then discuss a potential parallel, between type IIA and type IIB, where the D0 and D(-1) backgrounds are effectively accounted for by loops of higher-dimensional KK modes.
We conclude by discussing how certain higher-derivative corrections of the type IIB axio-dilaton sector can be written in a 12d covariant form, and the limitations of such an interpretation.

\subsection{Recap: Perturbative String Theory and Higher-Derivative Supergravity}
The perturbative expansion of string amplitudes is organised by two parameters. The Regge slope $\alpha' = l_s^2$ controls the derivative expansion, while the string coupling $g_s$ controls the worldsheet genus expansion.
An $n$-point amplitude is a sum over genus-$g$ worldsheet path integrals
\begin{equation}
    \mathcal A_n(\alpha') = \sum_{g=0}^\infty g_s^{2g-2} \mathcal A_n^{(g)}(\alpha').
\end{equation}
The tree-level 4-point function in the type IIA and type IIB string theories is the Virasoro-Shapiro amplitude \cite{Green:2012oqa}. 
For the symmetric traceless modes in the NSNS sector of the string, it takes, up to an irrelevant overall factor, the form \cite{Green:1999pv}
\begin{equation}
\begin{aligned}
\mathcal A_4
&=\frac{\alpha^{\prime 4}}{3\cdot 2^3 g_s^2}
K
\times\frac{64}{\alpha^{\prime 3}stu}
\frac{\Gamma[1-\tfrac{\alpha'}{4}s]\Gamma[1-\tfrac{\alpha'}{4}t]\Gamma[1-\tfrac{\alpha'}{4}u]}{\Gamma[1+\tfrac{\alpha'}{4}s]\Gamma[1+\tfrac{\alpha'}{4}t]\Gamma[1+\tfrac{\alpha'}{4}u]},\\
K &\equiv t_8^{m_1...m_8}t_8^{n_1...n_8}               
\prod_{r=1}^4\zeta^{(r)}_{m_{2r}n_{2r}}k^{(r)}_{m_{2r-1}}k^{(r)}_{n_{2r-1}},
\end{aligned}
\end{equation}
where $s,t,u$ are Mandelstam variables, and $\zeta^{(r)}_{mn}$ is the dimensionless polarisation of the symmetric traceless modes on a closed string.
They may also be interpreted as polarisations of the spacetime graviton $\kappa h^{(r)}_{mn}$ in $G_{mn}=\eta_{mn} + \kappa h_{mn}$.
The explicit form of $t_8 t_8$ can be found in \cite{Schwarz:1982jn,Policastro:2008hg,Liu:2025uqu}.
Using the relation
\begin{equation}
\ln\Gamma(1-x)-\ln\Gamma(1+x) = 2\gamma x + 2\sum_{m\ge0}\frac{\zeta(2m+3)}{2m+3}x^{2m+3},
\end{equation}
we find the expansion
\begin{equation}
\mathcal{A}_4 = \frac{\alpha^{\prime 4}}{3\cdot 2^3 g_s^2}
K
\left[ \frac{64}{\alpha^{\prime 3}stu} + 2 \zeta(3)\right]
 + O(\alpha^{\prime 5})
\end{equation}
evaluated in the string frame with $G_{mn} |_{\infty} = \eta_{mn}$.
From the relation $G_{mn} = e^{\Phi/2}g_{mn}$, we see that the Einstein frame metric is not asymptotically flat: $g_{mn} |_{\infty} = g_s^{-1/2}\eta_{mn}$.
To pass to the Einstein frame, we convert the implicit string frame contractions to those with the Einstein frame metric, which introduces factors of $g_s$.
The net effect is the following Einstein frame expansion
\begin{equation}
\mathcal{A}_4 \bigg|_{g_{mn}} = 
\frac{\alpha^{\prime 4}}{3\cdot 2^3 }
K
\times
\left[ \frac{64}{\alpha^{\prime 3}stu} + 2 \zeta(3)g_s^{-3/2}\right]
 + O(\alpha^{\prime 5}),
\label{eq:4-pt-virasoro-shapiro}
\end{equation}
where $\zeta_{mn}^{(r)}$ inside $K$ is now the polarisation of the Einstein frame graviton, and all contractions are performed with the asymptotic Einstein frame metric $g_s^{-1/2}\eta_{mn}$.

\begin{figure}[ht]
\centering
\begin{tikzpicture}[
    line cap=round,
    vrtx/.style={circle, fill=black, inner sep=1.3pt},
    leg/.style={thick},
    prop/.style={thick},
    op/.style={font=\large}
  ]
  % ---- s-channel ----
  \begin{scope}[xshift=0cm]
    \coordinate (sL) at (-0.4,0);
    \coordinate (sR) at (0.4,0);
    \draw[leg] (-1.1,0.7) -- (sL);
    \draw[leg] (-1.1,-0.7) -- (sL);
    \draw[prop] (sL) -- (sR);
    \draw[leg] (sR) -- (1.1,0.7);
    \draw[leg] (sR) -- (1.1,-0.7);
    \node[vrtx] at (sL) {};
    \node[vrtx] at (sR) {};
  \end{scope}
  \node[op] at (1.9,0) {$+$};
  % ---- t-channel ----
  \begin{scope}[xshift=3.8cm]
    \coordinate (tT) at (0,0.5);
    \coordinate (tB) at (0,-0.5);
    \draw[leg] (-1.1,0.7) -- (tT) -- (1.1,0.7);
    \draw[leg] (-1.1,-0.7) -- (tB) -- (1.1,-0.7);
    \draw[prop] (tT) -- (tB);
    \node[vrtx] at (tT) {};
    \node[vrtx] at (tB) {};
  \end{scope}
  \node[op] at (5.7,0) {$+$};
  % ---- u-channel ----
  \begin{scope}[xshift=7.6cm]
    \coordinate (uT) at (0,0.5);
    \coordinate (uB) at (0,-0.5);
    \draw[leg] (-1.1,0.7) -- (uT);
    \draw[leg] (-1.1,-0.7) -- (uB);
    \draw[prop] (uT) -- (uB);
    \draw[leg] (uT) -- (1.1,-0.7);
    \draw[leg] (uB) -- (1.1,0.7);
    \node[vrtx] at (uT) {};
    \node[vrtx] at (uB) {};
  \end{scope}
  \node[op, anchor=west] at (9.0,0) {$\sim \dfrac{1}{stu}$};
\end{tikzpicture}
\caption{The three graviton-exchange channels ($s$, $t$, and $u$, from left to right) that generate the nonlocal $1/stu$ pole of the four-graviton amplitude $\mathcal{A}_4$. Each arises from massless graviton exchange through the cubic vertices of the Einstein-Hilbert action.}
\label{fig:stu-channels}
\end{figure}

The polynomial $K$ is analytic in the momenta and arises from the expansion of $\frac{t_8 t_8 R^4}{16}$.
The nonlocal poles at $s, t, u = 0$ correspond to massless graviton exchange and reproduce the $s,t,u$-channel structure generated by the cubic graviton vertices of the Einstein-Hilbert action (Figure~\ref{fig:stu-channels}). The nonlocal $1/stu$ piece is therefore attributed to tree-level supergravity dynamics.
Indeed, it is accompanied by a single factor of $\alpha'$, which reflects its origin in two-derivative supergravity.

The next contribution in \eqref{eq:4-pt-virasoro-shapiro} appears at $O(\alpha^{\prime 4})$, which reveals an 8-derivative spacetime origin.
Its nonlinear completion is captured by $t_8t_8 R^4$:
\begin{equation}
\frac{\alpha^{\prime 4}}{3 \cdot 2^3} K 2 \zeta(3) g_s^{-3/2}
\to
\frac{\alpha^{\prime 4}}{3 \cdot 2^7} t_8t_8 R^4 (2 \zeta(3) g_s^{-3/2}) + O(\alpha^{\prime 5}).
\end{equation}

The leading correction in $g_s$ comes from genus-1 amplitudes \cite{Green:1999pv}, which are common to both type IIA and IIB theories and again carry the $t_8 t_8 R^4$ factor. 
Combining the genus-0 and genus-1 amplitudes and promoting $g_s$ to $\frac{1}{\tau_2}$, we obtain the local effective action
\begin{equation}
\mathcal{A}_4 = -4\pi^7 (t_8 t_8 R^4)\alpha^{\prime 4}
\left[
2 \zeta(3)\tau_2^{3/2}
+\frac{2\pi^2}{3}\tau_2^{-1/2}
\right]
+O(\alpha^{\prime 5}).
\label{eq:genus-0-genus-1-R4}
\end{equation}
Non-renormalisation theorem suggest that the perturbative corrections to $R^4$ terminate here at one loop \cite{Green:1997tv,Kehagias:1997cq}.

Non-perturbatively, the type IIB string path integral also requires summing over the D(-1) backgrounds \eqref{D-1 background}.
The singly and multiply charged D(-1) backgrounds give rise to non-perturbative $R^4$ corrections \cite{Green:1997tv,Kehagias:1997cq} of the form
\begin{equation}
\mathcal A_{D(-1)}\propto \alpha^{\prime 4}t_8t_8R^4\sum_{m,n\ge1}\left(\frac{m}{n^3}\right)^{1/2}
(e^{2\pi i mn\tau}+e^{-2\pi i mn\bar{\tau}})
\left(
1+\sum_{k=1}^\infty(4\pi mn\tau_2)^{-k}\frac{\Gamma[k-1/2]}{\Gamma[-k-1/2]k!}
\right).
\end{equation}
When combined with the genus-0 and genus-1 perturbative contributions, these D(-1) terms assemble into the modular-invariant, non-holomorphic Eisenstein series
\begin{equation}
\begin{aligned}
f_0(\tau,\bar\tau)
&=\sum_{(m,n)\neq(0,0)}\frac{\tau_2^{3/2}}{|m+n\tau|^3}\\
&=2\zeta(3)\tau_2^{3/2}+\frac{2\pi^2}{3}\tau_2^{-1/2}\\
&\quad+4\pi^{3/2}\sum_{m,n\ge1}\left(\frac{m}{n^3}\right)^{1/2}
(e^{2\pi i mn\tau}+e^{-2\pi i mn\bar{\tau}})
\left(
1+\sum_{k=1}^\infty(4\pi mn\tau_2)^{-k}\frac{\Gamma[k-1/2]}{\Gamma[-k-1/2]k!}
\right),
\end{aligned}
\label{eq:f0-eisenstein}
\end{equation}
thereby restoring the \SLZ{} duality symmetry, after the 2-derivative \SLR{} had been explicitly broken in \eqref{eq:genus-0-genus-1-R4}.
This is the weight-$\tfrac32$ non-holomorphic Eisenstein series $f_0 = E_{3/2}$. 
\footnote{We follow the convention that the $s$-th Eisenstein series is given by 
\begin{equation}
E_s(\tau,\bar\tau) \equiv \sum_{(m,n)\neq(0,0)} \frac{\tau_2^{s}}{|m+n\tau|^{2s}}.
\end{equation}}
Since type IIA has no D(-1) backgrounds, the restoration of \SLZ{} is unique to the type IIB theory.

\subsection{KK-modes and D-brane Backgrounds}
We saw above how summing over the D(-1) backgrounds completes the perturbative $R^4$ corrections into the Eisenstein series $f_0(\tau,\bar\tau)$, thereby restoring \SLZ{}.
Before turning to the 12d repackaging of the effective action, we note a closely related point: the supergravity KK-modes on a torus reproduce the effect of the D0 and D(-1) backgrounds in string path integrals.

We work in 12d with coordinates $(x^\mu,y^1,y^2)$ and compactify $y^1,y^2$ on a torus through the identifications
\begin{equation}
y^1\to y^1+2\pi R_T,\quad
(y^1,y^2)\to (y^1+2\pi R_T\tau_1,y^2+2\pi R_T\tau_2)
\end{equation}
where $R_T$ is an overall radius parameter setting the scale of the torus. A 12d scalar can then be Fourier expanded on $T^2$ as
\begin{equation}
\chi(x^\mu,y^1,y^2)
=\sum_{m,n}\chi_{m,n}(x^\mu) \times\exp
\left[
\frac{i}{R_T\tau_2}\left(m\tau_2y^1 + (n-m\tau_1)y^2\right)
\right].
\end{equation}
The massive modes are
\begin{equation}
[-\nabla^2_{10}-\nabla^2_2]\chi_{p,m,n}
=\left(p_{10}^2+M_{m,n}^2\right)\chi_{p,m,n},\quad
M^2_{m,n}
= \frac{m^2}{R_T^2}+\frac{(n-m\tau_1)^2}{R_T^2\tau_2^2}
=\frac{|n-m\tau|^2}{R_T^2\tau_2^2}.
\end{equation}
For 4-point amplitudes in 10d, we evaluate the contributions from loops of the infinite tower of massive KK-modes using the Schwinger proper-time representation
\cite{Green:1997as,Beccaria:2023hhi}
\begin{equation}
\mathcal{A}_4
\propto \sum_{n,m}\int_{\Lambda^{-2}}^\infty
\frac{d\lambda}{\lambda^{3/2}}
e^{-\lambda \frac{|n-m\tau|^2}{R_T^2\tau_2^2}}
P(s,t;\lambda),\quad
P(s,t;\lambda) = \int_{\Delta} d^3\rho  e^{-\lambda M(s,t;\rho)}
\label{4-point Schwinger parameter}
\end{equation}
where the Mandelstam variables $s,t,u$ are defined as usual, the Feynman parameters $\rho_i$ are integrated over the simplex $\Delta \equiv \{0\le\rho_1\le\rho_2\le\rho_3\le1\}$, and $M(s,t;\rho)$ is defined as
\begin{equation}
M(s,t;\rho)=s\rho_1\rho_2+t\rho_2\rho_3+u\rho_1\rho_3 + t(\rho_1-\rho_2).
\end{equation}
The local higher-derivative terms come from expanding \eqref{4-point Schwinger parameter} for small $s,t,u$.\footnote{The leading coefficient also follows from the shortcut of adding a spectator dimension to the known 11d-on-$T^2$ results \cite{Green:1997as}, but the direct route taken here keeps the dimension dependence explicit.}
Expanding the Feynman-parameter integral, we find
\begin{equation}
P(s,t;\lambda)
= \sum_{k=0}^\infty \frac{(-\lambda)^k}{k!}\int_{\Delta} d^3\rho  M(s,t;\rho)^k.
\end{equation}
Each power of $M$ carries two derivatives, so an order-$k$ term corresponds to a coupling of the form $D^{2k}R^4$.
The $R^4$ coupling of interest is the leading ($k=0$) term, for which $M$ drops out and $P$ reduces to the volume of the simplex $\Delta$,
\begin{equation}
P(0,0;\lambda) = \int_\Delta d^3\rho = \frac{1}{6}.
\end{equation}
Substituting this in the amplitude \eqref{4-point Schwinger parameter} gives
\begin{equation}
\mathcal{A}_4
\propto \sum_{(m,n)\neq(0,0)} \int_{\Lambda^{-2}}^\infty \frac{d\lambda}{\lambda^{3/2}}  e^{-\lambda M_{m,n}^2}.
\end{equation}
The massless mode $(m,n)=(0,0)$ is dropped, as it is infrared divergent.
The proper-time integral has an ultraviolet divergence at $\lambda\to0$, regulated by the cutoff $\Lambda$. 
Being local, it is absorbed into an $R^4$ counterterm.
Its renormalised finite part is fixed by the analytic continuation of $\int_0^\infty d\lambda \lambda^{s-1}e^{-\lambda M^2}=\Gamma(s)(M^2)^{-s}$.
We thus have
\begin{equation}
\mathcal{A}_4
\propto \frac{1}{R_T \sqrt{\tau_2}}\Gamma[-\tfrac{1}{2}] E_{-1/2}(\tau,\bar\tau),
\end{equation}
where we have substituted $M_{m,n}^2 = |n-m\tau|^2/(R_T^2\tau_2^2)$ and recognised the sum over $(m,n)\neq(0,0)$ as the Eisenstein series $E_{-1/2}(\tau,\bar\tau)$.
Using the functional relation
\begin{equation}
\pi^{-s}\Gamma(s)\zeta(2s)E_s(\tau,\bar\tau)=\pi^{-(1-s)}\Gamma(1-s)\zeta(2-2s)E_{1-s}(\tau,\bar\tau),
\end{equation}
we arrive at
\begin{equation}
\mathcal{A}_4 \propto R_T^{-1}\tau_2^{-1/2} f_0(\tau,\bar\tau).
\label{eq:KK-R4-coefficient}
\end{equation}
The prefactor $R_T^{-1}\tau_2^{-1/2}$ is itself modular invariant as $R_T$ transforms under \SLZ{}.
The $\tau$-dependence is carried by the modular-invariant Eisenstein series $f_0$ of \eqref{eq:f0-eisenstein}: a single KK-loop sum reproduces the series otherwise obtained by summing over the D(-1) backgrounds.
The KK tower on the torus therefore acts as a surrogate for the D(-1) background, in parallel with 10d type IIA supergravity, where the KK modes on $S^1$ produce the D0 background \cite{Green:1997as}.

We do not wish to overclaim.
The KK-loop reproduction of $f_0$ probes the established eleven-to-nine-dimensional relation \cite{Green:1997as}: it is compatible with a 12d-to-10d reading but does not establish one, since the $R^4$ coupling is not an intrinsically 10d quantity.
A genuinely 10d diagnostic requires an amplitude sensitive to all ten momenta, such as $\epsilon_{10}\epsilon_{10}R^4$, which first appears at five points.
One may attempt to evaluate KK-loop contributions to 5-point amplitudes using the 5-point Schwinger proper-time formula analogous to \eqref{4-point Schwinger parameter}, given in \cite{Minasian:2015bxa,Green:1999by,Liu:2022bfg}.

\subsection{12d Effective Corrections}
We now turn to the 12d interpretation of the type IIB higher-derivative corrections.
We begin by reviewing the known 12d interpretations of the type IIB effective action, using a ``bottom-up'' approach.
The type IIB supergravity in the axio-dilaton sector can be written as
\begin{equation}
2\kappa^2 S_{IIB,\tau}
= \int d^{10}x \sqrt{-g} \left(R - 2 P_m \overline{P}^m\right),
\label{eq:IIB-two-derivative}
\end{equation}
where $P_m \equiv \frac{i\nabla_m \tau}{2\tau_2}$.

We have previously discussed that for the D7 brane and the D(-1) instanton backgrounds, the axio-dilaton profile $\tau$ can be geometrised as the complex structure of a torus on which 12d gravity is compactified.
This proposal can be understood by noting that the 10d axio-dilaton action can be obtained from a certain 12d action on a specific metric ansatz, where the 12d metric is given by
\begin{equation}
\widehat{g}_{MN}=\begin{pmatrix}
g_{mn} & 0 \\
0 &\widehat{g}_{ab}
\end{pmatrix},\quad
\widehat{g}_{ab}
=\frac{1}{\tau_2}\begin{pmatrix}
1& \tau_1\\
\tau_1 & \tau_1^2+\tau_2^2
\end{pmatrix}.
\label{eq:F-theory-ansatze}
\end{equation}
One may adopt complex tangent frame coordinates $z, \overline{z}$, with vielbein $e_{\alpha}{}^a = \frac{1}{2\sqrt{\tau_2}}\begin{pmatrix}
    -\tau & 1 \\ -\overline{\tau} & 1
\end{pmatrix}$. 
The 12d metric in complex coordinates is then
$\widehat{g}_{zz} = \widehat{g}_{\overline{z}\overline{z}} = 0$, $\widehat{g}_{z\overline{z}} = 1/2$.
The nonvanishing components of the 12d Riemann tensor are (see also \cite{Liu:2025uqu})
\begin{equation}
\begin{gathered}
\widehat{R}_{mrns} = R_{mrns},\quad
\widehat{R}_{z\overline{z}z\overline{z}} = -\frac14 |P|^2,\quad
\widehat{R}_{mnz\overline{z}} = P_{[m}\overline{P}_{n]},\\
\widehat{R}_{mzn\overline{z}} = - \frac{P_m \overline{P}_n}{2},\quad
\widehat{R}_{mznz} = \frac{\nabla_m P_n}{2}.
\end{gathered}
\label{eq:12d-Riemann-complex}
\end{equation}
Substituting $P_m\overline{P}_n$ in \eqref{eq:IIB-two-derivative} with $\widehat{R}_{mzn\overline{z}}$, we find
\begin{equation}
\sqrt{-g}[R - 2 P_m \overline{P}^m]
=\sqrt{-\widehat{g}}[R + \widehat{R}^{mz}{}_{mz} + \widehat{R}^{m\overline{z}}{}_{m\overline{z}}]
=\sqrt{-\widehat{g}} \widehat{R},
\label{eq:2-derivative-index-splitting}
\end{equation}
which suggests that the F-theory ansatz \eqref{eq:F-theory-ansatze} relates the 10d axio-dilaton action to a 12d Einstein-Hilbert term.
Hence we may map solutions of 12d Einstein gravity, such as the pp-wave and the KK-monopole solutions, to type IIB geometries with nontrivial axio-dilaton profile.
This was discussed in section~\ref{sec:axio-dilaton-12d}.

It is natural to ask whether type IIB higher-derivative corrections (e.g. at 8-derivatives) can also be understood from a 12d perspective.
The answer does not appear to be a clear-cut yes or no, and the goal of the rest of this section is to clarify the current understanding.

\subsubsection{Type IIB at 4 points}

The 4-point 8-derivative effective action can be written \cite{Liu:2025uqu,Minasian:2015bxa,Policastro:2008hg} as
\begin{equation}
\begin{aligned}
\frac{L_{\text{4pt}}}{\beta  f_{0}(\tau,\overline{\tau})}
&= t_{8}t_{8} \biggl[R^4+6R^{2}\left(4|\delta\nabla P|^{2}+|\nabla G_{3}|^{2}\right)+24|\delta \nabla P|^{2}|\nabla G_{3}|^{2}\\
&+12R\left(\delta \nabla P(\nabla\overline{G}_{3})^{2}+\delta \nabla \overline{P}(\nabla{G}_{3})^{2}\right)\biggr]\\
&+1536 |\nabla_m P_n \nabla^m P^n|^2\\
&+O_{2}\left((|\nabla G_{3}|^{2})^{2}\right)
\label{eq:PTaction}
\end{aligned}
\end{equation}
where $\beta = \frac{\alpha^{\prime 3}}{3 \cdot 2^{12}}$ and $(\delta \nabla P)_{mrns} \equiv g_{[m[n} \nabla_{r]} P_{s]}$ is defined to have the same symmetries as the Riemann tensor, up to $\delta$-functions located at the sources of $\tau$.
We make an observation that the 3-form sector of \eqref{eq:PTaction} admits a factorisation that is compatible with a 12d interpretation, but further investigations are out of the scope of this paper.%
\footnote{
\label{footnote:3-form-factorisation}
The 3-form sector of \eqref{eq:PTaction} admits the factorisation
\begin{equation}
L_{\text{3-form}}
\propto
t_8t_8
6 |R \nabla G_3 + 2 \delta \nabla P \nabla \bar{G}_3|^2
+O_{2}\left((|\nabla G_{3}|^{2})^{2}\right).
\end{equation}
In the 12d perspective, $\delta \nabla P$ consists of linear combinations of the 12d Riemann tensor $\widehat{R}$, so schematically this factor is of the form $|\widehat{R} \widehat{F}_4|^2$ where $\widehat{F}_4$ is the 12d uplift of the 10d 3-form field strength $H_3, F_3$.
}

Restricting to the gravity + scalar sector of \eqref{eq:PTaction}, we have
\begin{equation}
\frac{L_{\text{4pt},\tau}}{\beta f_0(\tau,\overline{\tau})} = t_8t_8
(R^4 + 24 R^2 |\delta \nabla P|^2) + 1536 (\nabla_m P_n \nabla^m P^n)(\nabla_r \overline{P}_s \nabla^r \overline{P}^s).
\label{eq:4pt-gravity-axiodilaton}
\end{equation}
In the 12d perspective $\tau$ is embedded in the metric.
If certain axio-dilaton couplings in the 10d effective action admit a 12d interpretation, one should be able to rewrite such terms using 12d-covariant differential operators acting on the metric, i.e. the 12d Riemann tensor \eqref{eq:12d-Riemann-complex}. 
Once we have replaced the 10d fields and their derivatives with their 12d counterparts, we may reexamine the resulting expression, and see if the terms group into a natural pattern as a result of index splitting.
We carried out this exercise at the two-derivative level in \eqref{eq:2-derivative-index-splitting}. We now perform the same exercise at the 8-derivative level on \eqref{eq:4pt-gravity-axiodilaton}, as a ``bottom-up'' rederivation of the results of \cite{Minasian:2015bxa}.

Let us begin by expanding the $t_8t_8$ contractions.
The $t_8t_8$ contraction with tensors of symmetries of the Riemann tensor is known \cite{Policastro:2008hg} to be 
\begin{equation}
\begin{aligned}
\frac{t_8t_8 R^4}{3 \cdot 2^7}
&=  R_{pnqm} R^{prqs} R^{mt}{}_{ru} R^{nu}{}_{st} + \frac{1}{2}R_{mpnq} R^{qrps} R_{tsur} R^{untm} \\
& - \frac{1}{2}R_{mnrs} R^{rsnp} R_{pquv} R^{qmuv} - \frac{1}{4} R_{mnrs} R^{rs}{}_{pq} R^{qmuv} R^{np}{}_{uv}\\
& + \frac{1}{16} R_{mnrs} R^{mnpq} R_{pquv} R^{uvrs} + \frac{1}{32}(R_{mnrs} R^{mnrs})^2.
\end{aligned}
\label{eq:t8t8R4}
\end{equation}
Expanding the $t_8 t_8 R^2 |\delta \nabla P|^2$ sector, we find\footnote{
The computation is done with Cadabra, which agrees with the results of \cite{Minasian:2015bxa}. We have taken $f_0(\tau,\overline{\tau})$ to be $2\zeta(3) \tau_2^{3/2}$.
}
\begin{equation}
\begin{aligned}
\frac{1}{3 \cdot 2^{8}\zeta(3) \tau_2^{3/2}}\frac{L_{\text{4pt},\tau}}{\beta}
&=  \frac{t_8 t_8 R^4}{3 \cdot 2^7}
+ 4 \nabla_n P_m \nabla^r \overline{P}^s R^{mt}{}_{ru} R^{nu}{}_{st} \\
&+ 4 \nabla_m P_n \nabla^r \overline{P}^s R_{tsur} R^{untm}
+ 4 |\nabla_m P_n \nabla^m P^n|^2 \\
&- 4 \nabla_m P_n \nabla^n \overline{P}^r R_{rsuv} R^{smuv}
+ \frac{1}{2} \nabla_{m} P_{n} \nabla^{m}\overline{P}^{n}R^{rstu}R_{rstu} + O(R^2 P^3).
\end{aligned}
\end{equation}
The next step is to replace the 10d terms with their 12d counterparts. The only 10d terms that remain are $\nabla P$.
We note that, because the only 12d Riemann tensor linear in $P$ is $R_{mznz}$ and its conjugate, we have
\begin{equation}
\widehat{R}_{ambn} \widehat{R}^{a}{}_{r}{}^{b}{}_{s} = 2 \operatorname{Re}\left(\nabla_m P_n \nabla_r \overline{P}_s\right) + O(P^3).
\end{equation}
We will use $m, n, r, s, ...$ for 10d spacetime indices and $a, b, c, d, ...$ for 2d torus indices.
Substituting this as well as $t_8 t_8 R^4$ into \eqref{eq:4pt-gravity-axiodilaton}, we find
\begin{equation}
\begin{aligned}
\frac{1}{3 \cdot 2^{8}\zeta(3) \tau_2^{3/2}}\frac{L_{\text{4pt},\tau}}{\beta}
&=
R_{pnqm} R^{prqs} R^{mt}{}_{ru} R^{nu}{}_{st}
+ 2 \widehat{R}_{anbm} \widehat{R}^{arbs} R^{mt}{}_{ru} R^{nu}{}_{st}\\
& + \frac{1}{2}\Big[ R_{mpnq} R^{qrps} R_{tsur} R^{untm}
+ 4 \widehat{R}_{ambn} \widehat{R}^{bras} R_{tsur} R^{untm}\\
&\qquad + 2 \widehat{R}_{ambn} \widehat{R}^{bras} \widehat{R}_{csdr} \widehat{R}^{dncm}\Big]\\
& - \frac{1}{2}\left[ R_{mnrs} R^{rsnp} R_{pquv} R^{qmuv}
 + 4 \widehat{R}_{mban} \widehat{R}^{anbp} R_{pquv} R^{qmuv} \right]\\
& - \frac{1}{4} R_{mnrs} R^{rs}{}_{pq} R^{qmuv} R^{np}{}_{uv}\\
& + \frac{1}{16} R_{mnrs} R^{mnpq} R_{pquv} R^{uvrs}\\
& + \frac{1}{32}(R_{rstu} R^{rstu})
\left[R_{mnrs} R^{mnrs}+ 8 \widehat{R}_{ambn} \widehat{R}^{ambn}\right]+ O(\text{5-pt}).
\end{aligned}
\label{eq:4pt-gravity-axiodilaton-12d-replaced}
\end{equation}

In this expression we have grouped terms with the same index contraction structure together. 
We again see contraction patterns that could reveal a 12d covariant structure.
For example
\begin{equation}
\widehat{R}_{MNRS}\widehat{R}^{MNRS} = R_{mnrs} R^{mnrs} + 4 \widehat{R}_{ambn} \widehat{R}^{ambn} + O(P^3),
\end{equation}
the factor of 4 arises from the 4 ways to choose 2 torus legs in $\widehat{R}_{MNRS}$ that contribute to 4-pt amplitudes.
Hence the last term in \eqref{eq:4pt-gravity-axiodilaton-12d-replaced} is compatible with arising from a 12d contraction 
\begin{equation}
\frac{1}{32} (\widehat{R}_{MNRS} \widehat{R}^{MNRS})^2 \in t_8 t_8 \widehat{R}^4.
\end{equation}

Upon further inspection, one finds that every term in \eqref{eq:4pt-gravity-axiodilaton-12d-replaced} descends from a single 12d contraction, displayed below with the 12d contraction on the left and its 10d expansion (the pure 10d quartic followed by the mixed-leg terms) on the right:
\begin{equation}
\begin{aligned}
\widehat{R}_{PNQM} \widehat{R}^{PRQS} \widehat{R}^{MT}{}_{RU} \widehat{R}^{NU}{}_{ST}
&= R_{pnqm} R^{prqs} R^{mt}{}_{ru} R^{nu}{}_{st}\\
&\quad + 2 \widehat{R}_{anbm} \widehat{R}^{arbs} R^{mt}{}_{ru} R^{nu}{}_{st}+ \ldots\\
\widehat{R}_{MPNQ} \widehat{R}^{QRPS} \widehat{R}_{TSUR} \widehat{R}^{UNTM}
&= R_{mpnq} R^{qrps} R_{tsur} R^{untm}\\
&\quad + 4 \widehat{R}_{ambn} \widehat{R}^{bras} R_{tsur} R^{untm}\\
&\quad + 2 \widehat{R}_{ambn} \widehat{R}^{bras} \widehat{R}_{csdr} \widehat{R}^{dncm}+ \ldots\\
\widehat{R}_{MNRS} \widehat{R}^{RSNP} \widehat{R}_{PQUV} \widehat{R}^{QMUV}
&= R_{mnrs} R^{rsnp} R_{pquv} R^{qmuv}\\
&\quad + 4 \widehat{R}_{mban} \widehat{R}^{anbp} R_{pquv} R^{qmuv}+ \ldots\\
\widehat{R}_{MNRS} \widehat{R}^{RS}{}_{PQ} \widehat{R}^{QMUV} \widehat{R}^{NP}{}_{UV}
&= R_{mnrs} R^{rs}{}_{pq} R^{qmuv} R^{np}{}_{uv}+ \ldots\\
\widehat{R}_{MNRS} \widehat{R}^{MNPQ} \widehat{R}_{PQUV} \widehat{R}^{UVRS}
&= R_{mnrs} R^{mnpq} R_{pquv} R^{uvrs}+ \ldots\\
(\widehat{R}_{MNRS}\widehat{R}^{MNRS})^2
&= (R_{rstu} R^{rstu})
\left[R_{mnrs} R^{mnrs}+ 8 \widehat{R}_{ambn} \widehat{R}^{ambn}\right]+ \ldots
\end{aligned}
\end{equation}
where the ellipses denote terms with more than 4 fields in 10d, or terms that vanish when summed with the coefficients in \eqref{eq:t8t8R4}.
Thus, every term in \eqref{eq:4pt-gravity-axiodilaton-12d-replaced} has an origin from a 12d contraction $t_8 t_8 \widehat{R}^4$, upon substituting the specific F-theory ansatz \eqref{eq:F-theory-ansatze}.

The relation above was first noted in \cite{Minasian:2015bxa} in a ``top-down'' fashion, by making an informed guess of a 12d effective action, then showing that the resulting 10d action matches the known gravity and axio-dilaton amplitudes.
The ``bottom-up'' approach outlined here, despite requiring more exercises in index-relabelling and tensor identities, is perhaps more systematic, as it does not require one to make a 12d ansatz a priori, and the argument is purely based on covariance.
If there were to exist a 12d covariant expression that reproduces any 10d effective action, then after substituting 12d objects for 10d ones, one must find that the terms group into a more natural pattern arising from splitting 12d indices.
What would be a genuine obstruction to a 12d covariant interpretation is if certain terms cannot be written using covariant 12d objects at all.

\subsubsection{Type IIB at 5 points: Obstruction to 12d Covariance?}
\label{sec:5-pt-obstruction}

Recently, the type IIB 5-point 8-derivative effective action has been computed \cite{Liu:2025uqu,Claasen:2026owa}.
In the U(1)-preserving sector at 5 points, there exists a term, which realises \SLZ{}-invariance on its own, with the form (we drop all numerical factors)
\begin{equation}
L_{\text{5pt},\tau} \supset
\operatorname{Re}\left(R^{mnsr} \nabla_m \overline{P}_s \nabla_n \overline{P}_r P_q P^q\right).
\label{eq:12d-obstruction}
\end{equation}
Both $R^{mnsr}$ and $\nabla_m \overline{P}_s$ can be written as 12d covariant objects (see \eqref{eq:12d-Riemann-complex}).
However, while we may write $P_q \overline{P}^q$ using $\widehat{R}_{z\overline{z}z\overline{z}}$, there is no 12d covariant object that can be substituted for $P_q P^q$.
Building on the 5-point action of \cite{Liu:2025uqu}, we now study this term more carefully, and show that it constitutes a genuine obstruction to a 12d covariant interpretation of the type IIB effective action.

We had previously worked with, from \eqref{eq:12d-Riemann-complex}, the 12d Riemann tensor components for the complex tangent frame coordinates $\widehat{g}_{zz} = \widehat{g}_{\overline{z}\overline{z}} = 0$, $\widehat{g}_{z\overline{z}} = 1/2$. 
For the arguments we are about to make, we find it more convenient to work in real coordinates, where the torus metric is given by \eqref{eq:F-theory-ansatze}.
In real coordinates, the components of the 12d Riemann tensor $\widehat{R}$ are linear combinations of the basis structures $\nabla_m \nabla_n \tau_1, \nabla_m \nabla_n \tau_2, \nabla_m \tau_1 \nabla_n \tau_2, \dots$.
We list the components of $\widehat{R}_{ambn}$ in Table \ref{tab:12d-Riemann-components}.

\begin{table}[ht]
\centering
{\small
\setlength{\tabcolsep}{6pt}
\renewcommand{\arraystretch}{1.4}
\begin{tabular}{ccccccc}
\toprule
$\widehat{R}_{ambn}$
  & $\nabla_m\nabla_n\tau_1$
  & $\nabla_m\nabla_n\tau_2$
  & $\nabla_m\tau_1\nabla_n\tau_1$
  & $\nabla_m\tau_2\nabla_n\tau_2$
  & $\nabla_m\tau_1\nabla_n\tau_2$
  & $\nabla_m\tau_2\nabla_n\tau_1$ \\
\midrule
$\widehat{R}_{1m1n}$ & $0$ & $\frac{1}{2\tau_2^2}$ & $\frac{1}{4\tau_2^3}$ & $-\frac{3}{4\tau_2^3}$ & $0$ & $0$ \\
$\widehat{R}_{1m2n}$ & $-\frac{1}{2\tau_2}$ & $\frac{\tau_1}{2\tau_2^2}$ & $\frac{\tau_1}{4\tau_2^3}$ & $-\frac{3\tau_1}{4\tau_2^3}$ & $\frac{1}{4\tau_2^2}$ & $\frac{3}{4\tau_2^2}$ \\
$\widehat{R}_{2m1n}$ & $-\frac{1}{2\tau_2}$ & $\frac{\tau_1}{2\tau_2^2}$ & $\frac{\tau_1}{4\tau_2^3}$ & $-\frac{3\tau_1}{4\tau_2^3}$ & $\frac{3}{4\tau_2^2}$ & $\frac{1}{4\tau_2^2}$ \\
$\widehat{R}_{2m2n}$ & $-\frac{\tau_1}{\tau_2}$ & $\frac{\tau_1^2-\tau_2^2}{2\tau_2^2}$ & $\frac{\tau_1^2-3\tau_2^2}{4\tau_2^3}$ & $\frac{\tau_2^2-3\tau_1^2}{4\tau_2^3}$ & $\frac{\tau_1}{\tau_2^2}$ & $\frac{\tau_1}{\tau_2^2}$ \\
\bottomrule
\end{tabular}
}
\medskip
\caption{Nonvanishing mixed components $\widehat{R}_{ambn}$ of the 12d Riemann tensor in real torus coordinates, with torus indices $a,b\in\{1,2\}$ and spacetime indices $m,n$.}%
\label{tab:12d-Riemann-components}
\end{table}

As for the rest of the nonvanishing 12d Riemann tensor components, they are given by
\begin{equation}
\begin{aligned}
\widehat{R}_{mnab} 
&= 2\widehat{R}_{[a|m|b]n}\\
\widehat{R}_{1212}
&= -\frac{1}{2}g^{mn}g^{ab} \widehat{R}_{ambn}.
\end{aligned}
\end{equation}

The factor $P_q P^q$ in \eqref{eq:12d-obstruction} is a scalar quadratic in the first derivatives of $\tau$, with no second derivatives. 
A 12d-covariant expression for it must then be linear in the 12d Riemann tensor, since a product of two curvatures already carries four derivatives. 
For the ansatz \eqref{eq:F-theory-ansatze}, every nonvanishing component of $\widehat{R}$ is either the pure 10d Riemann tensor or, by the identities above, reduces to the mixed components $\widehat{R}_{ambn}$ of Table~\ref{tab:12d-Riemann-components}. 
The most general two-derivative scalar that depends on $\tau$ is therefore
\begin{equation}
S = c^{ab}  g^{mn} \widehat{R}_{ambn},
\end{equation}
where $c^{ab}(\tau)$ is an arbitrary symmetric tensor on the torus, built from $\widehat{g}^{ab}$. 
Reading off the table, $S$ is a linear combination of $\nabla^2\tau_1$, $\nabla^2\tau_2$ and the products 
$(\nabla\tau_1)^2$, $(\nabla\tau_2)^2$, $\nabla_m\tau_1\nabla^m\tau_2$, with $\tau$-dependent coefficients. 
Since $P_q P^q$ contains products of first derivatives only, the values of $c^{ab}$ must be chosen so the $\nabla^2\tau_1$ and $\nabla^2\tau_2$ pieces of $S$ cancel. 
This fixes $c^{ab}$ up to an overall factor, and the surviving combination is then forced to be
\begin{equation}
S \propto (\nabla\tau_1)^2 + (\nabla\tau_2)^2,
\end{equation}
with the two squares appearing with the \emph{same} sign and no cross term. 
By contrast,
\begin{equation}
P_m P^m = -\frac{1}{4\tau_2^2}
\left[
    (\nabla\tau_1)^2 - (\nabla\tau_2)^2 + 2i \nabla^m \tau_1 \nabla_m \tau_2
\right]
\label{eq:PP-real}
\end{equation}
carries the two squares with \emph{opposite} sign together with a nonvanishing cross term. 
No choice of $c^{ab}$ reproduces this structure, so $P_q P^q$, and with it the term \eqref{eq:12d-obstruction}, admits no 12d-covariant expression.

The deviation from the $t_8 t_8 \widehat{R}^4$ prediction at five points in the U(1)-preserving sector was already observed in \cite{Liu:2025uqu}. The analysis above shows that the term \eqref{eq:12d-obstruction} evades any covariant 12d contraction, not only the $t_8 t_8 \widehat{R}^4$ form.
It remains a possibility that higher-point amplitudes may contain additional terms that conspire with \eqref{eq:12d-obstruction} to support a 12d covariant uplift, e.g. by combining with \eqref{eq:12d-obstruction} to form a 12d covariant object.
To investigate this possibility, one needs to compute the 8-pt axio-dilaton effective action.\footnote{A single 12d Riemann tensor $\widehat{R}$ upon \eqref{eq:F-theory-ansatze} may reduce to, schematically, $\nabla P$ or a product of $P \bar{P}$ in 10d.
So a 12d structure of $\widehat{R}^4$ contributes to 4, 5, 6, 7 and 8-point amplitudes in 10d.}

\section{Concluding Remarks and Outlook}
\label{sec:outlook}

The 12d picture is most robust in the axio-dilaton sector (Section~\ref{sec:axio-dilaton-12d}), where the D7 and D(-1) backgrounds arise respectively as a 12d Kaluza-Klein monopole and pp-wave reduced on a torus, the type IIB \SLZ{} acts as large gauge transformations in 12d, and the reduction reproduces the type IIB Killing spinor equations.
Beyond this sector the interpretation is only partial (Section~\ref{sec:covariant-unification}).
The RR and NSNS form fields can be assembled into 12d objects, with the 10d self-duality of the 5-form realised as a 12d Hodge duality, but this is a covariant repackaging of the 10d theory on a two-torus: not all 12d degrees of freedom are excited, and the mismatch between the SO(8) and SO(10) little-group content shows that a full 12d uplift of the field content is unavailable.
On the D3 brane (Section~\ref{sec:d3-self-duality}), the worldvolume electromagnetic \SLZ{} duality is consistent only when accompanied by the bulk \SLZ{} transformation, tying the worldvolume and bulk dualities together \cite{Tseytlin:1996it}, though its 12d interpretation remains open.
The higher-derivative corrections finally make the limits of the interpretation precise (Section~\ref{sec:effective-actions}): at four points the 8-derivative effective action can be written in 12d-covariant form, but at five points a U(1)-preserving obstruction appears, whose fate at higher points remains open.

Taken together, the 12d interpretation is robust for the axio-dilaton sector but, beyond it, remains effective rather than fundamental.
This is consistent with the long-standing obstruction to a 10+2d supergravity: what emerges is a 12d-covariant organisation of the 10d theory rather than a dynamical 12d spacetime.

Two directions stand out for future work.
The first concerns the effective action: whether a more general 12d ansatz, going beyond the $t_8 t_8 \widehat{R}^4$ contraction, can accommodate the type IIB couplings, and whether the five-point obstruction persists at higher points. 
It is possible that, beginning at six points, further obstructive terms appear that conspire with the five-point obstruction (see discussion in Section~\ref{sec:5-pt-obstruction}) to restore a 12d-covariant form.
Confirming this would require deriving the axio-dilaton effective action up to eight points.
The alternative is that the obstruction is genuine, and the type IIB effective action cannot be written 12d-covariantly.
In addition, the 3-form sector of the 8-derivative effective action admits a factorisation that is compatible with a 12d interpretation (see Footnote~\ref{footnote:3-form-factorisation}), and may be worth further investigation.

The second concerns the role of \SLZ{} in the context of the AdS/CFT correspondence.
Recently, it was shown that an M2 brane wrapping a circle at the boundary of the $\mathrm{AdS}_4 \times S^7/\mathbb{Z}_k$ background reproduces, via a one-loop computation of its worldvolume effective action, the subleading $1/N$ corrections to Wilson-loop observables in the dual ABJM theory \cite{Giombi:2023vzu, Tseytlin:2024euk}.
This raises the question of whether an analogous construction exists for the $\mathrm{AdS}_5 \times S^5$ background, whose dual $\mathcal{N}=4$ SYM theory carries its own \SLZ{} duality.
Addressing it would require identifying the type IIB counterparts of the $\mathrm{AdS}_4 \times S^7/\mathbb{Z}_k$ geometry and the M2 brane, and in particular a 12d interpretation of the D3 brane that sources $\mathrm{AdS}_5 \times S^5$.
As discussed in Section~\ref{sec:d3-self-duality}, this is the hardest part of the 12d picture: the two worldvolume gauge degrees of freedom of the D3 have no direct interpretation as transverse 12d coordinates, and the one construction that does tie the D3 self-duality to a higher-dimensional geometry, the 6d $(2,0)$ theory on $T^2$, again passes through the 9d detour (see discussion at the end of Section~\ref{sec:d3-self-duality}).
A direct 12d realisation of the D3 worldvolume \SLZ{} would geometrise the $\mathcal{N}=4$ SYM coupling as the modulus of the reduction torus, in the same way the $(p,q)$ 7-branes geometrise the bulk \SLZ{} in Section~\ref{sec:axio-dilaton-12d}.

Recently, the 10d non-supersymmetric type 0 string theories have received renewed attention.
It has been argued that the type 0A theory arises as M-theory compactified on a ``figure-8'' circle $S^1 \wedge S^1$ \cite{Baykara:2026gem}, while the type 0B theory with gauge group $SO(16)^4$ has been suggested to arise as M-theory on $S^1 \wedge S^1 \times S^1 / \mathbb{Z}_2$ \cite{Altavista:2026evd}.
An extension of the web of dualities into the type 0 theories has also been explored \cite{Baykara:2026gem, Baykara:2026vdc}.
How a 12d perspective may fit into this picture remains an interesting open question.

\section*{Acknowledgements}

This paper is adapted from the author's MSc thesis, submitted to the Department of Physics, Imperial College London.
We would like to thank Arkady Tseytlin for their guidance and many valuable discussions.
We would also like to thank Amihay Hanany for their lectures on string theory at Imperial College London.

\appendix
\section{Dynkin Labels}
\label{sec:Dynkin Label Conventions}
Our convention for Dynkin labels is illustrated with the $B_4$ and $D_4$ diagrams
\begin{equation}
\begin{aligned}
[n_1, n_2, n_3, n_4]_{SO(9)} &= \dynkin[labels={n_1,n_2,n_3,n_4}, scale = 2]{B}{4},\\
[n_1, n_2, n_3, n_4]_{SO(8)} &= \dynkin[labels={n_1,n_2,n_3,n_4}, scale = 2]{D}{4}.
\end{aligned}
\end{equation}
With this convention, below we tabulate the common representations of the orthogonal groups.
\begin{center}
\begin{tabular}{p{0.35\textwidth} p{0.25\textwidth} p{0.25\textwidth}}
\toprule
\multicolumn{3}{c}{\textbf{Representations of } $SO(2n)$} \\
\midrule
\textbf{Dynkin Label} & \textbf{Dimension} & \textbf{Field} \\
\midrule
$[0,0\ldots 0]$                     & $1$                    & Scalar \\
$[1,0\ldots 0]$                     & $2n$                   & 1-form \\
$[0,1,0\ldots 0]$                  & $\binom{2n}{2}$        & 2-form \\
1 in $k$-th position                 & $\binom{2n}{k}$        & $k$-form \\
$[0\ldots 0,1,1]$                  & $\binom{2n}{n-1}$      & $(n-1)$-form \\
$[0\ldots 0,2,0] + [0\ldots 0,0,2]$
                                     & $\binom{2n}{n}$        & SD and ASD $n$-form \\
$[0\ldots 0,1,0]$                  & $2^{n-1}$            & LH Spinor \\
$[0\ldots 0,0,1]$                  & $2^{n-1}$            & RH Spinor \\
$[1,0\ldots 0,1,0]$               & $(2n-1)2^{n-1}$      & LH Gravitino \\
$[1,0\ldots 0,0,1]$               & $(2n-1)2^{n-1}$      & RH Gravitino \\
$[2,0\ldots 0]$                     & $n(2n+1)-1$            & Graviton \\
\bottomrule
\end{tabular}
\end{center}

\vspace{1em}

\begin{center}
\begin{tabular}{p{0.35\textwidth} p{0.25\textwidth} p{0.25\textwidth}}
\toprule
\multicolumn{3}{c}{\textbf{Representations of } $SO(2n+1)$} \\
\midrule
\textbf{Dynkin Label} & \textbf{Dimension} & \textbf{Field} \\
\midrule
$[0,0\ldots 0]$                     & $1$                    & Scalar \\
$[1,0\ldots 0]$                     & $2n+1$                 & 1-form \\
$[0,1,0\ldots 0]$                  & $\binom{2n+1}{2}$      & 2-form \\
1 in $k$-th position                 & $\binom{2n+1}{k}$      & $k$-form \\
$[0\ldots 0,0,2]$                  & $\binom{2n+1}{n}$      & $n$-form \\
$[0\ldots 0,0,1]$                  & $2^{n}$              & Spinor \\
$[1,0\ldots 0,1]$                  & $(2n)2^{n}$          & Gravitino \\
$[2,0\ldots 0]$                     & $n(2n+3)$              & Graviton \\
\bottomrule
\end{tabular}
\end{center}

\section{The Construction of the \texorpdfstring{\SLR{}}{SL(2,R)}/SO(2) Coset}
\label{sec:sl2r-coset}

We briefly review the construction of the $\frac{\SLR{}}{SO(2)}$ coset representative, which is important for the discussion of the \SLR{} symmetry and the interplay between \SLZ{} and U(1)-violation.

The coset representative of $\frac{\SLR{}}{SO(2)}$ is an \SLR{} matrix $\mathcal{V}$ with a local SO(2) symmetry.
Generically, $\dim \SLR{} = 3$, so $\mathcal{V}$ is parameterised by three real scalars: $\mathcal{V} = \mathcal{V}(\tau_1, \tau_2, \theta)$.
It is convenient to separate the SO(2) factor, parameterised by $\theta$, using the Iwasawa decomposition \cite{iwasawa1949}
\begin{equation}
\mathcal{V} =
\begin{pmatrix}
1 & 0 \\
\tau_1 & 1
\end{pmatrix}
\begin{pmatrix}
\frac{1}{\sqrt{\tau_2}} & 0\\
0 & \sqrt{\tau_2}
\end{pmatrix}
\begin{pmatrix}
\cos\theta & \sin\theta\\
-\sin\theta & \cos\theta
\end{pmatrix}.
\label{Iwasawa-decomposition}
\end{equation}
In this representation, \SLR{} acts on $\mathcal{V}$ from the left, with $\Lambda \equiv \begin{pmatrix}
d&c\\b&a\end{pmatrix} \in \SLR{}$.
The representative $\mathcal{V}$ by itself is not physical due to the SO(2) gauge redundancy.
However, one can construct $\mathcal{V}\mathcal{V}^T$, which transforms covariantly under \SLR{}, removes the SO(2) factor, and gives the physical metric for the scalar manifold, parameterised by $\tau_1, \tau_2$:
\begin{equation}
\mathcal{M} = \mathcal{V}\mathcal{V}^T = \frac{1}{\tau_2}\begin{pmatrix}
1 & \tau_1\\
\tau_1 & |\tau|^2
\end{pmatrix}.
\end{equation}

To construct the kinetic term for the scalars, one builds the sl(2, $\mathbb{R}$)-valued Maurer-Cartan 1-form $\mathcal{V}^{-1}\partial_m\mathcal{V}$, which decomposes into the non-compact and compact elements of the sl(2,$\mathbb{R}$) algebra: $\mathcal{V}^{-1}\partial_m\mathcal{V} = P_m + Q_m$, where $P_m$ projects onto the non-compact generators and $Q_m$ is the SO(2) connection.
The kinetic term for the scalars is then given by $\mathrm{tr}(P^m P_m)$, which is positive definite, as ensured by tracing over the non-compact elements of a Lie algebra.
Computing $P, Q$, with $SO(2)\cong U(1)$ complexified via $e^{i\theta} = \cos\theta + i\sin\theta$, one finds
\begin{equation}
P = P_1\begin{pmatrix}
    0&1\\1&0
\end{pmatrix}+P_2\begin{pmatrix}
    1&0\\0&-1
\end{pmatrix},\quad
P_1 + iP_2 = e^{2i\theta}\frac{\partial \tau}{2 \tau_2},\quad
Q = \partial \theta - \frac{1}{2}\frac{\partial \tau_1}{\tau_2}.
\end{equation}
So the vielbein $P$ transforms covariantly under the local U(1) with charge 2, and the kinetic term $\frac{|\partial \tau|^2}{2\tau_2^2}$ is U(1) invariant.
This U(1) is independent of the global \SLR{} transformation, and is a gauge redundancy reflecting the choice of the U(1) frame.

The parameter $\theta$ here is the non-dynamical U(1) phase.
We can remove its presence in the action by picking a gauge $\theta = 0$, leaving two real scalars $\tau_1, \tau_2$ that parameterise the 2d coset $\frac{\SLR{}}{U(1)}$ without redundancy.
However, general \SLR{} transformations do not preserve the $\theta = 0$ gauge, so a compensating U(1) transformation given by $\theta = -\arg(c\tau+d)$ is induced when $\Lambda$ acts on $\mathcal{V}$ from the left.
This is independent of the local U(1) symmetry, which may be violated, but is part of the global \SLZ{} symmetry, which should always be preserved.
This is another way to view the breaking of the U(1) symmetry: the original U(1) is the SO(2) factor in \SLR{}, so when \SLR{} is broken to \SLZ{}, the U(1) is also broken.

Besides $P$, the other bosonic field that transforms under U(1) is the 3-form field strength $G_3 = e^{\Phi/2}(F_3 - \tau H_3)$, which carries charge 1 under U(1).
When computing higher-derivative corrections, one wishes to keep track of the U(1) charge for possible U(1)-violating corrections. In that context it is convenient to work with the complexified vielbein $P$ and 3-form field strength $G_3$.
It is important to distinguish this U(1) from the compensating U(1) used to remove the redundant scalar $\theta$ in the coset representative $\mathcal{V}$. The compensating U(1) is an \SLZ{} transformation, under which the effective action should always be invariant. For example, the modular form $f(\tau, \bar\tau)$ also carries a U(1)-like charge under \SLZ{}, but it does not carry local U(1) charge, and hence may contribute to \SLZ{}-invariant but U(1)-violating corrections.

\section{Differential Geometry}

Here we present our conventions for differential geometry and form fields, as well as some useful identities that are used in the main text.

\subsection*{Conventions}
We adopt the following conventions and normalisations for form fields
\begin{equation}
\begin{aligned}
F &\equiv \frac{1}{p!}F_{\mu_1...\mu_p} dx^{\mu_1}...dx^{\mu_p},\\
F_{\mu_1...\mu_p} &\equiv F_{[\mu_1...\mu_p]},\\
|F|^2 &\equiv \frac{1}{p!}F_{\mu_1...\mu_p} F^{\mu_1...\mu_p}=\frac{1}{p!}F^2,\\
(d F)_{\mu_1...\mu_{p+1}} &= (p+1)\partial_{[\mu_1} F_{\mu_2...\mu_{p+1}]},\\
(F^{(p)} \wedge G^{(q)})_{\mu_1...\mu_p \nu_1 ... \nu_q}
&=\frac{(p+q)!}{p!q!}F^{(p)}_{[\mu_1...\mu_p}  G^{(q)}_{\nu_1 ... \nu_q]}.
\end{aligned}
\end{equation}
We will use $\epsilon_{\mu_1 ... \mu_p}$ to denote the Levi-Civita tensor and $\varepsilon_{\mu_1 ... \mu_p}$ to denote the flat-space fully antisymmetrised symbol,
\begin{equation}
\epsilon_{\mu_1...\mu_D} = \sqrt{|g|}\varepsilon_{\mu_1...\mu_D},\quad
\varepsilon_{0,1,...,(D-2),(D-1)} = 1.
\end{equation}
Following this convention, we have
\begin{equation}
\epsilon_{\mu_1...\mu_p \lambda_1 ...\lambda_{D-p}}
\epsilon^{\nu_1...\nu_p \lambda_1 ...\lambda_{D-p}}
=(-1)^{[t]} p!(D-p)!\delta^{\nu_1...\nu_p}_{\mu_1...\mu_p}
\end{equation}
as well as
\begin{equation}
(*A)_{\mu_1...\mu_{D-p}} = \frac{1}{p!}
\epsilon_{\mu_1...\mu_{D-p}}{}^{\nu_1...\nu_p} A_{\nu_1...\nu_p}.
\end{equation}

\subsection*{Useful form-field identities}

One can then show that
\begin{equation}
*(*A) = (-1)^{[t]}(-1)^{p(D-p)}A,
\end{equation}
\begin{equation}
(*F)\wedge *(*F)=(-1)^{[t]}F\wedge *F,\quad
|*F|^2 = (-1)^{[t]}|F|^2,
\end{equation}
as well as
\begin{equation}
\frac{1}{(D-p-1)!}(*F)_{\mu ...}(*F)_\nu{}^{...}
=(-1)^{[t]}\left[
    \frac{1}{p!}F^2 g_{\mu\nu} - \frac{1}{(p-1)!}F_{\mu...}F_\nu{}^{...}
\right]
\end{equation}
\begin{proof}
\begin{equation}
\begin{aligned}
\frac{1}{(D-p-1)!}(*F)_{\mu ...}(*F)_\nu{}^{...}
&=\frac{1}{(D-p-1)!}\frac{1}{p!p!}
\epsilon_{\mu \mu_1...\mu_p \lambda_1...\lambda_{D-p-1}}
\epsilon_{\nu} {}^{\nu_1...\nu_p} {}^{\lambda_1...\lambda_{D-p-1}}
F^{\mu_1...\mu_p}F_{\nu_1...\nu_p}\\
&=\frac{p+1}{p!}(-1)^{[t]} g_{\nu\rho} \delta^{[\rho \nu_1...\nu_p]}_{[\mu \mu_1...\mu_p]}F^{\mu_1...\mu_p}F_{\nu_1...\nu_p}
\end{aligned}
\end{equation}
The term $\delta^{[\rho \nu_1...\nu_p]}_{[\mu \mu_1...\mu_p]}$
has in total $(p+1)!(p+1)!$ terms and is thus normalised by such a number. Of those terms, $(p+1)p!p!$ give 
$\delta^{\rho}_{\mu} \delta^{\nu_1...\nu_p}_{\mu_1...\mu_p}$
equivalent and $(p+1)(p)p!p!$ give $
-\delta^{\rho}_{\mu_1}\delta^{\nu_1...\nu_p}_{ \mu ...\mu_p}$
equivalent, hence the relation provided.
\end{proof}
As a corollary, suppose $S_{MN}$ is defined by
\begin{equation}
S_{MN}\big|_{F}\equiv
\frac{1}{2}\left[
    \frac{1}{(n-1)!}F_{M...}F_N^{...}
    -\frac{n-1}{D-2}\left(\frac{1}{n!}F^2\right)g_{MN}
\right],
\end{equation}
then
\begin{equation}
S_{MN}|_{*F} = -(-1)^{[t]}S_{MN}|_{F}.
\end{equation}

\subsection*{Useful gravity identities}

Under dilatation, the Ricci tensor transforms as
\begin{equation}
R(\Lambda^2 g_{mn})
=\Lambda^{-2}[R-2(d-1)\nabla^2\ln\Lambda - (d-2)(d-1)(\nabla \ln \Lambda)^2].
\end{equation}
For a brane-like metric
\begin{equation}
ds_D^2 = H^{a}ds_{d}^2 + H^bds_{D-d}^2,
\end{equation}
and $H\equiv H(r)$ with $r\equiv \sqrt{y^my^m}$ is a harmonic function in $(D-d)$-dimensions, usually taken to be
\begin{equation}
H = 1 + \frac{Q}{r^{\tilde{d}}},\quad \tilde{d} = D-d-2,
\end{equation}
The Ricci tensor is found to be
\begin{equation}
\begin{aligned}
R_{\mu\nu}
&=-\eta_{\mu\nu}(H')^2H^{a-b-2}\left[\frac{a(ad+b\tilde d-2)}{4}\right]\\
R_{mn} 
&= \frac{(H')^2}{4H^2}\left[-b(ad+b\tilde{d}-2)\delta_{mn}-\frac{y^m y^n}{r^2}\left[ad(a-b)-(ad+b\tilde{d})(b+2)\right]\right]\\
&+\frac{H'}{rH}\frac{(ad+b\tilde{d})}{2}\left[-\delta_{mn} + \frac{y^m y^n}{r^2}(2+\tilde{d})\right].
\end{aligned}
\end{equation}
Usually one adopts the ansatz that $A_{n-1} \propto H$, hence $F_n \propto H'$.
Then the energy-momentum tensor will be exactly quadratic in $H'$ up to some powers of $H$.
This demands
\begin{equation}
b=-ad/\tilde{d},
\end{equation}
and we can write down a simplified Ricci tensor
\begin{equation}
\begin{aligned}
R_{\mu\nu} 
&= \frac{a}{2}(H')^2H^{a-b-2} \eta_{\mu\nu}
=\frac{a}{2}(H')^2H^{\frac{ad+(a-2)(D-d-2)}{(D-d-2)}} \eta_{\mu\nu},\\
R_{mn}
&=\frac{(H')^2}{4H^2}\left[2b\delta_{mn}-ad(a-b)\frac{y^my^n}{r^2}\right]
=-\frac{ad}{2(D-d-2)}\frac{(H')^2}{H^2}\left[
\delta_{mn}+\frac{a(D-2)}{2}\frac{y^m y^n}{r^2}
\right].
\end{aligned}
\end{equation}

\section{Catalog of Branes}
\label{sec:catalog-of-branes}
Here we list the brane-like solutions of type IIA, type IIB, and 11d M-theory.

\subsection*{11d}

The action for 11d supergravity is
\begin{equation}
2\kappa_{11}^2 S = 
\int d^{11}x\sqrt{-g}\left[
    R-\frac{1}{2}| F_4|^2
\right]
-\frac{1}{6}\int  C_3 \wedge  F_4 \wedge  F_4.
\end{equation}
It has the following brane solutions.
\paragraph{M2}
\begin{equation}
ds^2 = H^{-2/3}ds_{1,2}^2 + H^{1/3}ds_8^2,\quad
(C_3)_{\mu_0 \mu_1 \mu_2} = \varepsilon_{\mu_0 \mu_1 \mu_2}(H^{-1}-1).
\end{equation}

\paragraph{M5}
\begin{equation}
ds^2 = H^{-1/3}ds_{1,5}^2 + H^{2/3}ds_5^2,\quad
(F_4)_{m_1 m_2 m_3 m_4} = \varepsilon_{m_1 m_2 m_3 m_4 n} \partial_n H.
\end{equation}

\subsection*{10d IIA}

The action in the string frame is
\begin{equation}
2\kappa_{10}^2 S_{IIA} =
\int d^{10}x\sqrt{-G}\left[
    e^{-2\Phi}\left(R+4|\partial\Phi|^2-\frac{1}{2}|H_3|^2\right)
    -\frac{1}{2}|F_2|^2-\frac{1}{4}|\tilde{F}_4|^2
\right]
-\frac{1}{2}\int B_2 \wedge F_4 \wedge F_4,
\end{equation}
where $H_3 = dB_2$, $F_2 = dC_1$, $F_4 = dC_3$, and $\tilde{F}_4 = F_4 - C_1 \wedge H_3$.
We use the convention $g_s = 1$ by shifting $\Phi \to \Phi - \langle\Phi\rangle$, where $e^{\langle\Phi\rangle} = g_s$.
The Einstein frame action is
\begin{equation}
\begin{aligned}
2\kappa_{10}^2 S_{IIA}
&= \int d^{10}x\sqrt{-g}\Bigg[
    R - \frac{1}{2}|\partial\Phi|^2 - \frac{1}{2}e^{-\Phi}|H_3|^2 \\
    &\quad - \frac{1}{2}e^{3\Phi/2}|F_2|^2 - \frac{1}{4}e^{\Phi/2}|\tilde{F}_4|^2
\Bigg]
- \frac{1}{2}\int B_2 \wedge F_4 \wedge F_4.
\end{aligned}
\end{equation}
It has the following brane solutions, with the metrics given in the Einstein frame.

\paragraph{D0}
\begin{equation}
\begin{aligned}
ds^2 &= H^{-7/8}ds_1^2 + H^{1/8}ds_9^2,\\
e^{\Phi} &= H^{3/4},\quad
(C_1)_0 = H^{-1}-1.
\end{aligned}
\end{equation}

\paragraph{F1}
\begin{equation}
\begin{aligned}
ds^2 
&= H^{-3/4}ds_{1,1}^2 + H^{1/4}ds_8^2,\\
e^{\Phi}
&= H^{-1/2},\quad
(H_3)_{\mu_0\mu_1 m} = \varepsilon_{\mu_0 \mu_1} \partial_m H^{-1}.
\end{aligned}
\end{equation}

\paragraph{D2}
\begin{equation}
\begin{aligned}
ds^2
&= H^{-5/8}ds_{1,2}^2 + H^{3/8}ds_7^2,\\
e^{\Phi}
&= H^{1/4},\quad
C_{012} = H^{-1}-1.
\end{aligned}
\end{equation}

\paragraph{D4}
\begin{equation}
\begin{aligned}
ds^2
&= H^{-3/8}ds_{1,4}^2 + H^{5/8}ds_5^2,\\
e^{\Phi}
&= H^{-1/4},\quad
C_{01234} = H^{-1}-1.
\end{aligned}
\end{equation}

\paragraph{NS5}
\begin{equation}
\begin{aligned}
ds^2 
&= H^{-1/4}ds_{1,5}^2 + H^{3/4}ds_4^2,\\
e^{\Phi}
&= H^{1/2},\quad H_3 = *_4 dH.
\end{aligned}
\end{equation}

\paragraph{D6}
\begin{equation}
\begin{aligned}
ds^2 
&= H^{-1/8}ds_{1,6}^2 + H^{7/8}ds_3^2,\\
e^{\Phi}
&= H^{-3/4},\quad
C_{0123456} = H^{-1}-1.
\end{aligned}
\end{equation}

\subsection*{10d IIB}

We use the convention $g_s = 1$.
The action in the string frame is
\begin{equation}
\begin{aligned}
2\kappa_{10}^2 S_{IIB} =
&\int d^{10}x\sqrt{-G}\left[
    e^{-2\Phi}\left(R+4|\partial\Phi|^2-\frac{1}{2}|H_3|^2\right)
    -\frac{1}{2}\left(|F_1|^2+|\tilde{F}_3|^2+\frac{1}{2}|\tilde{F}_5|^2\right)
\right]\\
&- \frac{1}{2}\int C_4\wedge H_3\wedge F_3,
\end{aligned}
\end{equation}
where $H_3 = dB_2$, $F_1 = dC_0$, $F_3 = dC_2$, $F_5 = dC_4$, $\tilde{F}_3 = F_3 - C_0 \wedge H_3$, and $\tilde{F}_5 = F_5 - \frac{1}{2}C_2 \wedge H_3 + \frac{1}{2}B_2 \wedge F_3$, with self-duality $\tilde{F}_5 = *\tilde{F}_5$.

In the Einstein frame it takes the form
\begin{equation}
\begin{aligned}
2\kappa_{10}^2S_{\mathrm{IIB}}
&=\int d^{10}x\sqrt{-g}
\Bigg[\left(R-\frac{\partial\bar\tau\partial\tau}{2\tau_2^2}\right)-\frac{1}{2}\left(e^{-\Phi}|H_3|^2+e^{\Phi}|\tilde F_3|^2\right)-\frac{1}{4}|\tilde F_5|^2\Bigg]\\
&\quad-\frac{1}{2}\int C_4\wedge H_3\wedge F_3.
\label{einstein frame}
\end{aligned}
\end{equation}

It has the following brane solutions, with the metrics given in the Einstein frame. The exception is the D(-1) instanton, whose metric we give in the string frame (denoted $ds_{(S)}^2$), since its Einstein-frame metric is flat.

\paragraph{D(-1)}
\begin{equation}
\begin{aligned}
ds_{(S)}^2 = H^{1/2}ds_{10}^2,\quad
e^{\Phi} = H,\quad
C_0 = H^{-1}-1.
\end{aligned}
\end{equation}

\paragraph{F1}
\begin{equation}
\begin{aligned}
ds^2
&= H^{-3/4}ds_{1,1}^2 + H^{1/4}ds_8^2,\\
e^{\Phi}
&= H^{-1/2},\quad
(H_3)_{\mu_0\mu_1 m} = \varepsilon_{\mu_0 \mu_1} \partial_m H^{-1}.
\end{aligned}
\end{equation}

\paragraph{D1}
\begin{equation}
\begin{aligned}
ds^2 
&= H^{-3/4}ds_{1,1}^2 + H^{1/4}ds_8^2,\\
e^{\Phi}
&= H^{1/2},\quad
(F_3)_{\mu_0\mu_1 m} = \varepsilon_{\mu_0 \mu_1} \partial_m H^{-1}.
\end{aligned}
\end{equation}

\paragraph{D3}

\begin{equation}
\begin{aligned}
ds^2 
&= H^{-1/2}ds_{1,3}^2 + H^{1/2}ds_6^2,\\
e^\Phi
&= \mathrm{const},\\
(F_5)_{\mu_0\mu_1\mu_2\mu_3 m} &= \varepsilon_{\mu_0\mu_1\mu_2\mu_3}\partial_m H^{-1},\quad
(F_5)_{m_1 m_2 m_3 m_4 m_5} = \varepsilon_{m_1m_2m_3m_4m_5 n}\partial_n H.
\end{aligned}
\end{equation}

\paragraph{NS5}
\begin{equation}
\begin{aligned}
ds^2 
&= H^{-1/4}ds_{1,5}^2 + H^{3/4}ds_4^2,\\
e^{\Phi}
&= H^{1/2},\quad H_3 = *_4 dH.
\end{aligned}
\end{equation}

\paragraph{D5}
\begin{equation}
\begin{aligned}
ds^2 
&= H^{-1/4}ds_{1,5}^2 + H^{3/4}ds_4^2,\\
e^{\Phi}
&= H^{-1/2},\quad F_3 = *_4 dH.
\end{aligned}
\end{equation}
\section{Reduction of the 12d Form-Field Sector}
\label{sec:form-field-reduction}
This appendix collects the wedge-product identities underlying the covariant 12d repackaging of the form fields in Section~\ref{sec:a-covariant-unification-in-12d}.
Throughout, the torus is oriented so that $\int_{T_2}dz_1\wedge dz_2 = -\mathrm{Vol}(T_2)$, as fixed in Section~\ref{sec:a-covariant-unification-in-12d}.

\subsection*{The four-form wedge product}
With $\widehat C_3 = B_2\wedge dz_1 + C_2\wedge dz_2$ and $\widehat F_4 = H_3\wedge dz_1 + F_3\wedge dz_2$, any term carrying a repeated torus leg vanishes, so
\begin{equation}
\widehat C_3\wedge\widehat F_4 = B_2\wedge dz_1\wedge F_3\wedge dz_2 + C_2\wedge dz_2\wedge H_3\wedge dz_1 .
\end{equation}
Anticommuting the torus one-forms through the odd-degree 10d forms, $dz_1\wedge F_3 = -F_3\wedge dz_1$ and $dz_2\wedge H_3 = -H_3\wedge dz_2$, and then using $dz_2\wedge dz_1 = -dz_1\wedge dz_2$, gives
\begin{equation}
\widehat C_3\wedge\widehat F_4 = \big(C_2\wedge H_3 - B_2\wedge F_3\big)\wedge dz_1\wedge dz_2 .
\label{eq:C3F4-wedge}
\end{equation}

\subsection*{The composite seven-form}
With $\widehat F_7 = F_5\wedge dz_1\wedge dz_2$ and \eqref{eq:C3F4-wedge}, the composite seven-form of Section~\ref{sec:a-covariant-unification-in-12d} is
\begin{equation}
\widetilde{\widehat F}_7 \equiv \widehat F_7 - \tfrac12\widehat C_3\wedge\widehat F_4
= \Big(F_5 - \tfrac12 C_2\wedge H_3 + \tfrac12 B_2\wedge F_3\Big)\wedge dz_1\wedge dz_2
= \tilde F_5\wedge dz_1\wedge dz_2 ,
\end{equation}
so its torus bracket is exactly the 10d composite $\tilde F_5 = F_5 - \tfrac12 C_2\wedge H_3 + \tfrac12 B_2\wedge F_3$.
The relative sign is fixed by this requirement: the opposite choice $\widehat F_7 + \tfrac12\widehat C_3\wedge\widehat F_4$ would reverse the two correction terms relative to $\tilde F_5$ and spoil the contraction $|\widetilde{\widehat F}_7|^2 = \det(\widehat g^{ab})|\tilde F_5|^2$.

\subsection*{The Chern-Simons coefficient}
A four-form wedged with itself retains only the cross term,
\begin{equation}
\widehat F_4\wedge\widehat F_4 = 2(H_3\wedge dz_1)\wedge(F_3\wedge dz_2) = -2H_3\wedge F_3\wedge dz_1\wedge dz_2 ,
\end{equation}
so with $\widehat C_4 = C_4$,
\begin{equation}
\widehat C_4\wedge\widehat F_4\wedge\widehat F_4 = -2C_4\wedge H_3\wedge F_3\wedge dz_1\wedge dz_2 .
\end{equation}
Integrating over the torus with the chosen orientation $\int_{T_2}dz_1\wedge dz_2 = -\mathrm{Vol}(T_2)$,
\begin{equation}
\frac12\int_{T_2\times\mathbb R^{1,9}}\widehat C_4\wedge\widehat F_4\wedge\widehat F_4
= \mathrm{Vol}(T_2)\int_{\mathbb R^{1,9}} C_4\wedge H_3\wedge F_3 ,
\end{equation}
reproducing the display identity of Section~\ref{sec:a-covariant-unification-in-12d}.
The factor of two from $\widehat F_4\wedge\widehat F_4$ is why the action prefactor in \eqref{12d action} is $-1/(8\kappa_{12}^2)$ rather than $-1/(4\kappa_{12}^2)$: substituting this identity into $-\tfrac{1}{8\kappa_{12}^2}\int\widehat C_4\wedge\widehat F_4\wedge\widehat F_4$ returns the type IIB Chern-Simons term $-\tfrac{1}{4\kappa^2}\int C_4\wedge H_3\wedge F_3$.

\bibliographystyle{JHEP}
\bibliography{references}

\end{document}